\documentclass[showpacs,preprintnumbers,amsmath,amssymb]{revtex4}
\usepackage{subfigure}
\usepackage{graphicx}

\begin{document}
\preprint{APS/123-QED}

\title{The extended, relativistic hyperon star model.}

\author{I. Bednarek}
\email{bednarek@us.edu.pl}
 \altaffiliation[]{}
\author{M.Keska }%
 \email{markeslio@pf.pl}
\affiliation{}%

\author{R. Manka}
\email{manka@us.edu.pl}
 \homepage{http://www.cto.us.edu.pl/~manka}
\affiliation{
 Department of Astrophysics and Cosmology, Institute of Physics, \\
 University of Silesia, Uniwersytecka 4, PL-40-007 Katowice, Poland
}%

\date{\today}% It is always \today, today,

\begin{abstract}
In this paper an equation of state of neutron star matter which includes strange baryons in the framework of Zimanyi and Moszkowski (ZM) model
has been obtained. We concentrate on the effects of the
isospin dependence of the equation of state constructing for the appropriate choices of parameters the hyperons star model.
Numerous neutron star models show that the  appearance of hyperons
is  connected with the increasing density in neutron
star interiors.  Various studies have indicated that the inclusion of $\delta$ meson mainly
affects the symmetry energy and through this the chemical composition of a neutron star. As the effective nucleon mass
contributes to hadron chemical potentials it alters the chemical composition of the star. In the result the obtained model of the star not only
excludes large population of hadrons but also does not reduce significantly lepton contents in the star interior.
\end{abstract}

\maketitle

%#######################################################################
\section{Introduction}
Compact stars which are known from observations can be classified
into two distinctive groups. The first one is exemplified by white
dwarfs and the second by neutron stars. Neutron stars themselves
are identify with pulsars and compact X-ray sources \cite{weber}.
At the core of a neutron star the density of matter ranges from a
few times of the density of normal nuclear matter to about an
order of magnitude higher. Thus various exotic forms of matter
such as hyperons or quark-hadron mix phase are expected to emerge
in the interior of a neutron star \cite{weber}. The appearance of
these additional degrees of freedom and their impact not only on
neutron stars but on proto-neutron stars structure and evolution
as well has been the subject of extensive studies \cite{glen}.
Properties of matter at such extreme densities are of particular
importance in determining forms of equations of state relevant for
neutron stars and in examining their global parameters
\cite{bema}. Theoretical description of hadronic systems should be
performed with the use of quantum chromodynamics (QCD) as it is
the fundamental theory of strong interactions. However, at the
hadronic energy scale where the observed experimentally degrees of
freedom are not quarks but hadrons  the direct description of
nuclei in terms of QCD become inadequate. Other alternative
approach has to be formulated one of which is quantum
hadrodynamics (QHD) \cite{walecka} giving quantitative description
of the nuclear many body problem. QHD is a relativistic quantum
field theory in which nuclear matter description in terms of
baryons and mesons is provided. The original model (QHD-I)
contains nucleons interacting through the exchange of simulating
medium range attraction $\sigma$ meson and $\omega$ meson
responsible for short range repulsion. Extension (QHD-II) of this
theory \cite{bog77,bodmer,gmuca} includes also the isovector meson
$\rho $. Nonlinear terms in the scalar and vector fields were
added in order to get the correct value of the compressibility of
nuclear matter and the proper density dependence in the vector
self-energy. The variation of nucleon properties in nuclear medium
is the key problem in nuclear physics. In order to incorporate
quark degrees of freedom in the analysis of nuclear many-body
system Guichon \cite{qmc1} provides a quark-meson coupling model
(QMC).  The extension of the QMC theory namely the quark mean
field model (QMF), describing a nucleon with the use of the
constituent quark  model, has been successfully applied to study
the properties of both nuclear matter and finite nuclei. The model
considered in this paper is an alternative version of the Walecka
approach with enlarged meson sector. In the interior of neutron
stars the density of matter could exceed normal nuclear matter
density up to a few times, in such high density regime nucleon
Fermi energies exceed the value of hyperon masses and thus the new
hadronic degrees of freedom are expected to emerge. The higher the
density the more various hadronic species are expected to
populate. The onset of hyperon formation depends on the
hyperon-nucleon and hyperon-hyperon interactions. Hyperons can be
formed both in leptonic and baryonic processes. Several relevant
strong interaction processes proceed and establish the hyperon
population in the neutron star matter. Neutron star models are
constructed at different levels of complexity starting from the
most elementary one which assumes that neutrons are the only
component. The more sophisticated version is formulated under the
assumption that the neutron star matter  has to obey the
constrains of charge neutrality and $\beta$ equilibrium. Thus the
model considered describes  high isospin asymmetric matter and it
has to be extended by the inclusion of isovector-scalar  meson
$a_{0}(980)$ ($\delta $ meson)\cite{delta}. For the sake of
completness additional nonlinear vector meson interactions are
included. When strange hadrons are taken into account
uncertainties which are present in the description of nuclear
matter are intensified due to the incompleteness of the available
experimental data. The standard approach does not reproduce the
strongly attractive hyperon-hyperon interaction seen in double
$\Lambda$ hypernuclei. In order to construct a proper model which
do include hyperons the effects of hyperon-hyperon interactions
have to be taken into account. These interactions are simulated
via (hidden) strange meson exchange: scalar meson $f_0 (975)$
($\sigma^*$ meson) and vector meson $\phi (1020)$ ($\phi $ meson)
and influence the form of the equation of state and neutron stars
properties. \newline The solution of the presented model is gained
with the mean field approximation in which meson fields are
replaced by their expectation values. The parameters used are
adjusted in the limiting density range around saturation density
$\rho_0$ and in this density range give very good description in
finite nuclei. However, incorporation of this theory to higher
density require an extrapolation which in turn leads to some
uncertainties and suffers of several shortcomings. The standard
TM1 parameter set for high density range reveals an instability of
neutron star matter which is connected with the appearance of
negative nucleon effective mass due to the presence of hyperons.
The Zimanyi-Moszkowski (ZM) \cite{zima} model in which the Yukawa
type interaction $g_{sN}\varphi$ is replaced by the derivative one
$(g_{sN}\varphi/M_N)\bar{\psi}_N\gamma_{\nu}\partial^{\nu}\psi$
exemplifies an alternative version of the Walecka model which
improves the behaviour of the nucleon effective masses. It also
influences the value of the incompressibility $K$ of neutron star
matter. The derivative coupling effectively introduces the density
dependence of the scalar and vector coupling constants. Knowing
the form of the equation of state (EOS) is the decisive factor in
determining properties of neutron stars such as: central density,
mass-radius relation, crust extent or the moment of inertia.\\ The
essential goal of this paper is to obtain within the described
above model the equation of state for the neutron star matter on
the basis of  calculations carried out for asymmetric nuclear
matter in the relativistic mean field approach (RMF) and to
compare the obtained results with ones that are relevant for high
density calculations - namely with a quark star. The inclusion of
$\delta$ meson affects the neutron stars chemical composition
changing the proton fraction which in turn affects the cooling
mechanism. If the proton fraction is higher than the critical
value of about $Y_p \sim 0.11$ \cite{lat} the direct URCA
processes can proceed and this enhances the rate of neutron star
cooling. Whether the proton fraction can exceed the critical value
and at what density it occurs depends on the model. The proton
fraction is almost entirely determined by the isospin-dependent
part of the EOS thus the inclusion of $\delta$ meson and nonlinear
vector meson interactions  influence the neutron star structure
and properties.\newline This paper is organized as follows.
Section 2 outlines the extended model with derivative coupling
including hyperons and additional mesons, together with the
collected equations of motions. Their solutions enable the
construction of the equation of state . In section 3  the
equilibrium conditions leading to the relevant hyperon star
composition are presented together with the chosen values of
parameters. Section 4 contains numerical results and the
discussion of their influence on neutron stars properties.
%#####################################################################
\section{The model}
For the description of  properties of the infinite nuclear matter
with nonzero strangeness the relevant parts of the SU(3) structure
have been involved. Due to the parity conservation the appearance
of pseudoscalar mesons is forbidden. The Lagrangian function for
the system can be written as a sum of a  baryonic part including
the full octet of baryons together with  baryon-meson interaction
terms, a mesonic part containing  additional interactions between
mesons which mathematically express themselves as supplementary,
nonlinear terms in the Lagrangian function, and a free leptonic
part
\begin{equation} \label{lag}
{\mathcal{L}}={\mathcal{L}_{BM}}+{\mathcal{L}_{M}} + {\mathcal{L}_{L}}.
%UWAGA\label{lg:sum}
\end{equation}
The interacting baryons are described by the Lagrangian function
${\mathcal{L}_{BM}}$ which is given by
\begin{equation}
{\mathcal{L}_{BM}}=\sum_{B}(1+\frac{g_{sB}\sigma
+g_{\sigma^*B}\sigma^*+I_{3B}g_{\delta
B}\tau^a\delta^a}{M_B})\bar{\psi}_Bi\gamma^{\mu}D_{\mu}\psi_B-\sum_B\overline{\psi}_BM_B\psi_B.
\end{equation}
where  the spinor $\Psi_B^T =(\psi_N,\psi_{\Lambda},\psi_{\Sigma},\psi_{\Xi})$ is composed of the following
isomultiplets \cite{glen},\cite{len}:
\[\begin{array}{cc}
\displaystyle\Psi_N={\psi_p \choose \psi_n}, &
\displaystyle\Psi_{\Lambda}=\psi_{\Lambda},\\ \\
\displaystyle\Psi_{\Sigma}=\left( \begin{array}{c}\psi_{\Sigma^+}\\ \psi_{\Sigma^0} \\ \psi_{\Sigma^-} \end{array}\right),&
\displaystyle\Psi_{\Xi}={\psi_{\Xi^0} \choose \psi_{\Xi^-}}.
\end{array}\]
The covariant derivative $D_{\mu}$ is defined as
\begin{equation}
D_{\mu}=\partial_{\mu}+ig_{\omega B}\omega_{\mu}+ig_{\phi B}\phi_{\mu}+ig_{\rho B}I_{3B}\tau^ab^a_{\mu}.
\end{equation}
The model considered represents an alternative version of the
Walecka model in which the Yukawa interaction term is replaced by
the derivative coupling one.  This Zimanyi-Moszkowski \cite{zima}
method allows to solve the problem of too low effective nucleon
masses achieved in the original approach. Rescaling the baryon
field in a way proposed by Zimanyi and Moszkowski \cite{zima} the
modified Lagrange function for interacting baryons is obtained
\begin{equation}
{\mathcal{L}_{B}}=-\sum_B\bar{\psi}_Bi\gamma^{\mu}D_{\mu}\psi
-\sum_B\left(1+\frac{g_{sB}\sigma
+g_{\sigma^*B}\sigma^*+I_{3B}g_{\delta
B}\tau^a\delta^a}{M_B}\right)^{-1}\overline{\psi}_BM_B\psi_B.
\end{equation}
The mesonic part of the Lagrangian function (\ref{lag}) is given by
\begin{eqnarray}
{\mathcal{L}_{M}} & = & \frac{1}{2}\partial _{\mu }\sigma \partial ^{\mu }\sigma -U(\sigma )+ \frac{1}{2}\partial _{\mu }\sigma^* \partial ^{\mu }\sigma^*- \frac{1}{2}m^{2}_{\sigma^*}\sigma^{*2}+\frac{1}{2}\partial _{\mu }\delta^a \partial ^{\mu }\delta^a -\frac{1}{2}m^{2}_{\delta}\delta ^{a}\delta^{a} \nonumber \\
& + &\frac{1}{2}m^{2}_{\phi}\phi_{\mu}\phi ^{\mu} -\frac{1}{4}\phi_{\mu \nu}\phi^{\mu \nu}
-\frac{1}{4}\Omega _{\mu \nu }\Omega ^{\mu \nu }+\frac{1}{2}m_{\omega }^2\omega _{\mu }\omega ^{\mu }+ (g_{\rho}g_{\omega})^2\Lambda_vb_{\mu}^ab^{a\mu}\omega_{\mu}\omega^{\mu} \nonumber \\
& &+(g_{\rho}g_{s})^2\Lambda_4b_{\mu}^ab^{a\mu}\varphi^2 -
\frac{1}{4}R_{\mu \nu }^{a}R^{a\mu \nu }+\frac{1}{2}m_{\rho}
^2b^{a}_{\mu }b^{a\mu }+\frac{1}{4}c_{3}(\omega _{\mu }\omega
^{\mu })^{2}+\frac{1}{4}\zeta (b^a_{\mu }b ^{a\mu })^{2}.
\label{lag1}
\end{eqnarray}
The field tensors \( R_{\mu \nu }^{a} \), \( \Omega _{\mu \nu } \),
\( \phi_{\mu \nu } \) are defined as
\begin{equation}
R_{\mu \nu }^{a}=\partial _{\mu }b^{a}_{\nu }-\partial _{\nu }b^{a}_{\mu }+g_{\rho }\varepsilon _{abc}b_{\mu }^{b}b_{\nu }^{c},
\end{equation}
\begin{equation}
\Omega _{\mu \nu }=\partial _{\mu }\omega _{\nu }-\partial _{\nu }\omega _{\mu },\hspace{2cm}\phi_{\mu \nu }=\partial _{\mu }\phi _{\nu }-\partial _{\nu }\phi _{\mu }.
\end{equation}
The potential function \( U(\sigma ) \) possesses the well-known
polynomial form introduced by Boguta and Bodmer \cite{bodmer}
\begin{equation}
U(\sigma )=\frac{1}{2}m^{2}_{s}\sigma ^{2}+\frac{1}{3}g_{3}\sigma ^{3}+\frac{1}{4}g_{4}\sigma ^{4}.
\end{equation}
The baryon mass is denoted by $M_B$ whereas \( m_{i} \) $(i=
\sigma ,\omega ,\rho ,\sigma^*,\phi ,\delta )$ are masses assigned
to the meson fields and they are taken at their physical values.
The parameters entering the Lagrangian function are the coupling
constants \( g_{\omega B} \), \( g_{\rho B} \), \( g_{s B} \),
$g_{\delta B}$, $g_{\sigma^*B}$, $g_{\phi B}$ for meson fields and
the self-interacting coupling constants \( g_{3} \), \( g_{4} \),
\( c_{3} \), $\zeta$. These parameters are adjusted to reproduce
the  bulk properties of the symmetric nuclear matter at
equilibrium. They are collected in Table 2. As the $\delta$ meson
field and nonlinear vector meson interactions carry isospin they
contribute to the symmetry energy $E_s$ of the system. This
imposes constrains on $g_{\rho}$ and $g_{\delta}$ $\Lambda_4$
$\Lambda_v$ and $\zeta$ parameters. They are adjusted to get the
experimental value of the symmetry energy $E_s$ which is equal $32
\pm 4$ MeV . Thus the parameter $g_{\rho}$ has been redefined by
comparison with the one in the original TM1 parameter set.\\ The
presence of hyperons demands additional coupling constants (they
are collected in Table 3) which have been established with the use
of experimental data and theoretical
analysis.\\
The inclusion of hyperons improve our understanding of neutron
stars at higher densities. There is common belief  that at
suitable densities a quark star is  more stable than the neutron
one. Thus it is interesting to compare the properties of hyperon
stars with those obtained in the framework of QMF model for a
quark star \cite{rmib}. Now the Lagrangian function possesses the
following form
\begin{eqnarray}
{\mathcal{L}_{QMF}}=\overline{q}(i\gamma ^{\mu }D_{\mu }-m_{c})q+\frac{1}{2}\partial _{\mu }\varphi _{a}\partial ^{\mu }\varphi _{a}-\frac{1}{2}m_{s}^{2}\varphi _{a}\varphi _{a} &  & \\
-\frac{1}{4}F_{\mu \nu }^{a}F^{a,\mu \nu }+\frac{1}{2}m_{v}^{2}V_{\mu }^{a}V^{a\mu } &  & \label{qlef}
\end{eqnarray}
where $q$ denotes a quark field with three flavors, $u$, $d$ and
$s$, and three colors. The construction of the model mimics the
relativistic mean field theory, where the scalar $\sigma $ and the
vector meson $\omega $ fields do not couple  with nucleons but
directly with quarks. The quark mass has to change from its bare
current mass due to the coupling to the \textbf{$\sigma $} meson.
In the framework of QMF model with the use of  the TM1 parameter
set quarks constituent masses are $m_{c,u}=m_{c,d}=367.61$ MeV and
$m_{c,s}=504.1$ MeV whereas  the bag constant is chosen at the
level of \textbf{$B^{1/4}=154.5$} MeV. There are no significant
differences between the presented above Lagrangian function
(\ref{qlef}) and the Lagrangian function (\ref{lag1}). They differ
in the baryonic sectors whereas the mesonic sectors for both
models are nearly the same. There are only differences in coupling
constants: $g_{\sigma q}= g_{\sigma B}/3, g_{\omega q }=g_{\omega
B }/3,
 g_{\rho q }=g_{\rho B}, g_{\delta q }=g_{\delta B} $.
We restrict ourselves to the isospin SU(2)
unbroken symmetric case, $m_{c}^{u}=m_{c}^{d}$, so \begin{equation}
m_{c}=m_{c,f}\delta _{f,f'}=\left(\begin{array}{ccc}
 m_{c,u} &  & \\
  & m_{c,d} & \\
  &  & m_{c,s}\end{array}
\right)\end{equation}
Generators of the U(3) algebra $\lambda ^{a}=\{\lambda ^{0}=\sqrt{2/3}I,\, \lambda ^{i}\}$
(where $I$ is an identity matrix, $\lambda ^{i}$are Gell-Mann matrices
SU(3) algebra) obeying $Tr(\lambda ^{a}\lambda ^{b})=2\delta _{ab}$.
Restricting only to $U(2)\times U(1)$ subalgebra ($a=\{0,1,2,3,8\}$)
case the simplest version of the QMF theory can be obtained. Defining the new
base with $\tau ^{a},\, \, a=\{0,1,2,3,4\}$ as \begin{equation}
\tau ^{0}=\left(\begin{array}{ccc}
 1 & 0 & 0\\
 0 & 1 & 0\\
 0 & 0 & 0\end{array}
\right)\, \, \, \tau ^{i}=\left(\begin{array}{cc}
 \sigma ^{i} & 0\\
 0 & 0\end{array}
\right)\, \, \, for\, \, \, i=\{1,2,3\},\, \, \tau ^{4}=\sqrt{2}\left(\begin{array}{ccc}
 0 & 0 & 0\\
 0 & 0 & 0\\
 0 & 0 & 1\end{array}
\right)\end{equation} the meson fields may be decomposed as
follows\[ \varphi =\varphi _{a}\tau ^{a}=\sigma \tau ^{0}+\delta
_{i}\tau ^{i}+\sigma ^{*}\tau ^{4}\] and \[ V_{\mu }=V_{\mu
}^{a}\lambda ^{a}=\omega _{\mu }\tau ^{0}+\phi_{\mu }\tau
^{4}+b_{i}\tau ^{i}\] Here the QMF model is enlarged by the
inclusion of the isovector $\delta $ meson.  Similarly to nucleons
it splits $u$ and $d$ masses. Both $\delta $ and $\rho $ mesons
may be neglected in the case of the symmetric nuclear matter.
\vspace*{1.5cm} \centerline{Baryon and meson
parameters}\vspace*{0.15in}
\begin{center}
\begin{tabular}{l|c c c c c c c} \hline
Particle & m [MeV] & J & I & S & Y & $I_{3}$ & Q \\ \hline
n & 939.566 & 1/2 & 1/2 & 0 & 1 & -1/2 & 0 \\
p & 938.272 & 1/2 & 1/2 & 0 & 1 & 1/2 & 1 \\
$ \Lambda$ & 1115.63 & 1/2 & 0 & -1 & 0 & 0 & 0 \\
$\Sigma^{-}$ & 1197.43 & 1/2 & 1 & -1 & 0 & -1 & -1 \\
$\Sigma^{0}$ & 1192.55 & 1/2 & 1 & -1 & 0 & 0 & 0 \\
$\Sigma^{+}$ & 1189.37 & 1/2 & 1 & -1 & 0 & 1 & 1 \\
$\Xi^{0}$ & 1314.9 & 1/2 & 1/2 & -2 & -1 & 1/2 & 0 \\
$\Xi^{-}$ & 1321.3 & 1/2 & 1/2 & -2 & -1 & -1/2 & -1 \\ \hline
$\sigma$ & 550 & 0 & 0 & 0 & 0 & 0 & 0 \\
$\omega$ & 783 & 1 & 0 & 0 & 0 & 0 & 0 \\
$\rho$ & 770 & 1 & 1 & 0 & 0 & -1,0,1 & -1,0,1 \\
$\delta$ & 980 & 0 & 1 & 0 & 0 & -1,0,1 & -1,0,1  \\
$\sigma^{\ast}$ & 975 & 0 & 0 & 0 & 0 & 0 & 0 \\
$\phi$ & 1020 & 1 & 0 & 0 & 0 & 0 & 0     \\ \hline
\end{tabular}
\end{center}
\newpage\centerline{Table 2} \vspace*{0.15in} \centerline{Parameter
sets used in this paper}\vspace*{0.15in}
\begin{center}
\begin{tabular}{c|c|c|c|c} \hline
Parameter & TM1 & ZM & ZM + $\delta$& ZM + $\delta$ + nonlinear terms\\ \hline M & 938 MeV & 938 MeV & 938 MeV & 938 MeV\\ $c_{3}$ & 71.3075 & 0 & 0 & 0 \\
$g_{3}$ & 7.23 $fm^{-1}$ & 0 & 0 & 0\\
$g_{4}$ & 0.6183 & 0 & 0 & 0\\
$g_{\omega N}$ & 12.6239 & 6.671 & 6.671 & 6.671\\
$g_{\sigma N}$ & 10.0289 & 7.8449 & 7.8449 & 7.8449\\
$g_{\rho N}$ & 9.2644 & 8.9 & 9.5 & 9.5\\
$g_{\delta N}$ & 0 & 0 & 3.1 & 3.1\\
$g_{\sigma^{\ast}N}$ & 0 & 0 & 0 & 0 \\
$g_{\phi N}$ & 0 & 0 & 0 & 0 \\
$\Lambda_{\textsc{v}}$ & 0 & 0 & 0 & 0.008 \\
$\Lambda_{4}$ & 0 & 0 & 0 & 0.001 \\
$\zeta$ & 0 & 0 & 0 & 0.5 \\ \hline
\end{tabular}
\end{center}
\vspace*{0.15in}\centerline{Table 3}
\vspace*{0.15in}\centerline{Hyperon-meson
couplings}\vspace*{0.15in}
\begin{center}
\begin{tabular}{c|c|c|c}\hline
Parameter & Value & Parameter & Value \\ \hline
$g_{\Lambda\sigma}$ & 0.5207 $g_{\sigma N}$ & $g_{\Lambda\sigma^{\ast}}$ & 0.5815 $g_{\sigma N}$\\
$g_{\Lambda\rho}$  & 0 & $g_{\Lambda\delta}$ &  0  \\
$g_{\Lambda\omega}$ & $\frac{2}{3}$ $g_{\omega N}$ & $g_{\Lambda\phi}$ & -$\frac{\sqrt{2}}{3}$ $g_{\omega N}$ \\
$g_{\Sigma\sigma}$ & 0.1565 $g_{\sigma N}$ &
$g_{\Sigma\sigma^{\ast}}$ & 0 \\ $g_{\Sigma\rho}$ & 2 $g_{\rho N}$
& $g_{\Sigma\delta}$ & 2 $g_{\delta N}$ \\ $g_{\Sigma\omega}$ &
$\frac{2}{3}$ $g_{\omega N}$ & $g_{\Sigma\phi}$ &
-$\frac{\sqrt{2}}{3}$ $g_{\omega N}$ \\ $g_{\Xi\sigma}$ & 0.2786 $g_{\sigma N}$ & $g_{\Xi\sigma^{\ast}}$ & 0.5815 $g_{\sigma N}$  \\
$g_{\Xi\rho}$& $g_{\rho N}$ & $g_{\Xi\delta}$ & $g_{\delta N}$ \\
$g_{\Xi\omega}$ & $\frac{1}{3}$ $g_{\omega N}$  & $g_{\Xi\phi}$ &
-$\frac{2\sqrt{2}}{3}$ $g_{\omega N}$ \\
\hline
\end{tabular}
\end{center}
\newpage
As it is usually assumed in quantum hadrodynamics
the mean field approximation is adopted and for the ground state of homogeneous infinite matter
quantum fields operators are replaced by their classical
expectation values. Thus one can separated mesonic fields into classical mean field values and quantum fluctuations
which are not included in the ground state:
\begin{eqnarray}
\sigma & = &\overline{\sigma} + s \\  \nonumber
\sigma^* & = &\overline{\sigma}^* + s^* \\  \nonumber
\delta^a& = & \overline{\delta}^a+ d\delta^{3a}  \\  \nonumber
\phi_{\mu} & = &\overline{\phi}_{\mu} + f_0\delta_{\mu 0} \\   \nonumber
\omega_{\mu}& = &\overline{\omega}_{\mu}+ w_{0}\delta_{\mu 0} \\   \nonumber
b_{\mu}^a & = & \overline{b}^a_{\mu}+r_{0}\delta_{\mu 0}\delta^{3a}
\end{eqnarray}
The derivative terms are neglected and if one assume homogenous
and isotropic infinite matter only time-like components of the
vector mesons will survive. The field equations derived from the
Lagrange function at the mean field level are
\begin{equation}
m_s^2s+g_3s^2+g_4s^2 -2(g_{\rho B}g_{sB})^2\Lambda_4r_0^2s=\sum_Bg_{sB}M^2_{eff,k_F}S(M_{eff,B},k_{F,B})
\end{equation}
\begin{equation}
m_{\omega}^2w_{0}+ c_3w_{0}^3+2(g_{\rho B}g_{\omega B})^2\Lambda_vr_0^2w_0=\sum_Bg_{\omega B}n_B
\end{equation}
\begin{equation}
m_{\rho}^2r_0+2(g_{\rho B}g_{sB})^2\Lambda_4r_0s^2+2(g_{\rho
B}g_{\omega B})^2\Lambda_vr_0w_0^2 =\sum_Bg_{\rho B}I_{3B}n_B
\end{equation}
\begin{equation}
m_{\delta}^2d^3=\sum_Bg_{\delta B}I_{3B}S(M_{eff,B},k_{F,B})
\end{equation}
\begin{equation}
m_{\sigma*}^2s^*=\sum_Bg_{\sigma^*B}M^2_{eff,k_F}S(M_{eff,B},k_{F,B})
\end{equation}
\begin{equation}
m_{\phi}^2f_0=\sum_Bg_{\phi B}n_B.
\end{equation}
The function $S(M_{eff,B},k_{F,B})$ is expressed with the use of an integral
\begin{equation}
S(M_{eff,B},k_{F,B})=\frac{2J_B+1}{2\pi^2}\int_0^{k_{FB}}\frac{M_{B,eff}}{\sqrt{k^2+M_{B,eff}}}k^2dk
\end{equation}
where $J_B$ and $I_{3B}$ are the spin and isospin projection of
baryon $B$, $k_{F,B}$ is the Fermi momentum of species $B$, $n_B$
denote the baryon  number density. The Dirac equation for baryons
that is obtained from the Lagrangian function has the following
form
\begin{equation}
(i\gamma ^{\mu }\partial_{\mu }-M_{B,eff}-g_{\omega
B}\gamma^{0}\omega_{0}-g_{\phi
B}\gamma^{0}f_{0}-\frac{1}{2}g_{\rho B}\gamma^{0}\tau^3r_{0})\psi
=0
\end{equation}
with $M_{B,eff}$ being the effective nucleon mass
generated by the nucleon and scalar fields interactions and is defined as
\begin{equation}
M_{B,eff}=\frac{M_B}{1+(g_{sB}s+g_{\sigma^*B}s^*+I_{3B}g_{\delta_B}d)/M_B}
\end{equation}
%#####################################################################
\section{The equilibrium conditions and composition of matter}
In the high density regime in neutron star interiors when the
Fermi energy of nucleons exceeds the hyperon masses additional
hadronic states are produced. The onset of hyperon formation depends on the attractive
hyperon-nucleon interaction. The higher the density the more various hadronic species
are expected to populate. They can be formed both in leptonic and baryonic processes.
In the last one the strong interaction process such as
\begin{equation}
n+n \rightarrow n+\Lambda
\end{equation}
proceeds.
There are other relevant strong  reactions that establish the hadron population in neutron star
matter e.g.:
\begin{equation}
\Lambda +n \rightarrow \Sigma^- +p \hspace{5mm} \Lambda +\Lambda \rightarrow \Xi^{-} + p
\end{equation}
The comparison of weak interaction time scales ($10^{-10}$ s) and
the time scale connected with the lifetime of a relevant star
indicate that there is a difference between the matter in high
energy collisions which is constrained by the isospin symmetry and
the strangeness conservation whereas neutron star matter by the
charge neutrality and the generalized $\beta$-equilibrium. Thus
realistic neutron star models describe electrically neutral high
density matter with no strangeness conservation being in $\beta $
equilibrium. The last condition implies the presence of leptons.
Mathematically it is expressed by adding the Lagrangian of free
leptons to the lagrangian function (\ref{lag})
\begin{equation}
\mathcal{L}_{L}=\sum_{l=e,\mu}\overline{\psi}_{l}(i\gamma ^{\mu }\partial_{\mu }-m_{l})\psi_{l}.
\end{equation}
Neutrinos are neglected here since they leak out from a neutron star, whose
energy diminishes at the same time.
After electron chemical potential $\mu_e$ reaches the value equal to the
muon mass, muons start to appear. Equilibrium with respect
to the reaction
\begin{equation}
e^{-}\leftrightarrow \mu^{-}+\nu_{e}+\bar{\nu}_{\mu}
\end{equation}
is assured when $\mu_{\mu} = \mu_e$ (setting $\mu_{\nu_e}=\mu_{\bar{\nu}_{\mu}}=0$).
The appearance of muons reduces number of  electrons and affects also the proton fraction.
The introduction of the asymmetry parameter $f_a$  which describes the relative neutron excess defined as
\begin{equation}
f_a=\frac{n_n-n_p}{n_N}
\end{equation}
allows to study the symmetry properties of the system. The
equilibrium conditions between  baryonic and leptonic species
which are present in the neutron star matter lead to the following
relations between their chemical potentials and constrain the
species fraction in the star interior
\begin{eqnarray}
\mu_p=\mu_{\Sigma^+}=\mu_n-\mu_e  \hspace{10mm}\mu_{\Lambda}=\mu_{\Sigma^0}=\mu_{\Xi^0}=\mu_n  \\ \nonumber
\mu_{\Sigma^-}=\mu_{\Xi^-}=\mu_{n}+\mu_{e} \hspace{10mm}  \mu_{\mu}=\mu_{e}
\end{eqnarray}
where the baryonic chemical potential is given by the relation
\begin{equation}
\mu_i=\sqrt{k^2_{F,B}+M^2_{B,eff}}+g_{\omega B}\omega_{0}+g_{\phi
B}f_{0}+I_{3B}g_{\rho B}r_0
\end{equation}
Similarly to the asymmetry parameter $f_a$ a parameter $f_s$ which
specify the strangeness content in the system and is strictly
connected with the appearance of particular hyperon species in the
model has been introduced.
\begin{equation}
f_s=\frac{n_{\Lambda}+n_{\Sigma}+2n_{\Xi}}{n_{\Lambda}+n_{\Sigma}+n_{\Xi}+n_{N}}.
\end{equation}
The RMF theory as an effective one requires the knowledge of
coupling constants. There are several selected parameterizations
of the theory strictly connected with the assumption that the
exchange of scalar, isoscalar-vector, vector mesons and two
hidden-strangeness scalar and vector mesons  are responsible for
interactions between specific constituents of the matter.
Particular sets of coupling constants are associated with the
description of nucleon-nucleon, hyperon-nucleon and
hyperon-hyperon interactions and  are indispensable for the the
construction of the  equation of state which in turn is applied
for determining a neutron star properties. The enlarged Walecka
model uses the TM1 parameter set (Table 2)\cite{tm1}. More
realistic description of a neutron star requires taking into
consideration not only the interior region of a neutron star but
also remaining layers, namely the inner and outer crust and
surface layers. The composite EOS constructed by adding Bonn
\cite{bonn} and Negele-Vautherin (NV) \cite{nv} equations of state
(describing the inner crust) to the TM1 one allows to calculate
the neutron star structure for the entire neutron star density
span. The parameters describing the nucleon-nucleon interactions
are created in order to reproduced the properties of the symmetric
nuclear matter at saturation such as the binding energy, symmetry
energy and the incompressibility.  Other parameters are related to
the hyperon-nucleon interactions. The scalar meson coupling to
hyperons can be calculated from the potential depth of the hyperon
in the saturated nuclear matter
\begin{equation}
U^{N}_{Y}(\rho_{0})=-g_{sY}\varphi +g_{\omega Y}\omega.
\end{equation}
The vector coupling constants for hyperons are determined from
$SU(3)$ symmetry as \cite{lamsig}
\begin{eqnarray}
\frac{1}{2}g_{\omega \Lambda}=\frac{1}{2}g_{\omega
\Sigma}=g_{\omega \Xi}=\frac{1}{3}g_{\omega N}  \\ \nonumber
\frac{1}{2}g_{\rho \Sigma}=g_{\rho \Xi}=g_{\rho N}; g_{\rho
\Lambda}=0  \\ \nonumber 2g_{\phi \Lambda}=2g_{\phi
\Sigma}=g_{\phi \Xi}=\frac{2\sqrt{2}}{3}g_{\omega N}.
\end{eqnarray}
The single $\Lambda$ potential in nuclear matter is well
determined as (\cite{ksi}) $-U^{N}_{\Lambda}(\rho_0)=27-28$ MeV.
Recent analysis of atomic data \cite{bart,gal2} indicate for
repulsive $\Sigma$ potential in the interior of nuclei
$U^{N}_{\Sigma}(\rho_0)=30$ MeV thus $\Sigma$ hyperons do not
appear in the bulk matter calculations. The interpretation of the
$\Xi$ hyperons data gives the value of the potential well depth
$-U^{N}_{\Xi}(\rho_0)=18$ MeV. The experimental data concerning
hyperon-hyperon  \cite{dhsf,mdg} interactions are extremely scare.
Analysis of events which can be interpreted as the creation of
$\Lambda \Lambda$ hypernuclei allows to determine the well depths
of hyperon in hyperon matter
\begin{equation}
-U_h(\rho_{0})=40 MeV
\end{equation}
\\
Having specified parameters of the model the equation of state can be calculated.
This has been done with the use of
the energy-momentum
tensor $T_{\mu\nu}$ defined as
\begin{equation}
T_{\mu \nu}=2\frac{\partial \mathcal{L}}{\partial g^{\mu \nu}}-g_{\mu \nu}\mathcal{ L}.
\end{equation}
It allows to calculate the pressure $P$ and energy density
$\varepsilon$ of the system. The pressure $P$ is related to the
statistical average of the trace of the spatial component $T_{ij}$
of the energy-momentum tensor \( P=\frac{1}{3}<T_{ii}> \), whereas
the energy density $\varepsilon$ equals \(<T_{00}> \). Thus the
complete form of the equation of state includes contributions
coming from meson, fermion and baryon fields and finally one can
get the following equations for the energy density $\varepsilon $
and pressure $P$
\begin{eqnarray}
\varepsilon & = &
\frac{1}{2}m_{\rho}^2(r_{0})^2+\frac{1}{2}m_{\delta
}d^2+\frac{1}{2}m_{\phi }f_0^2+\frac{1}{2}m_{\sigma^* }(s^*)^2\\
\nonumber & + & \frac{1}{2}m_{\omega
}^{2}w_{0}^{2}+\frac{3}{4}c_3w_{0}^{4} +U(s)+3\Lambda_{v}
g_{\omega}^{2}g_{\rho}^{2}r_{0}^{2}w_{0}^{2} +
\Lambda_{4}g_{\sigma}^{2}g_{\rho}^{2}r_{0}^{2}\varphi^{2}+
\frac{1}{8}\zeta g_{\rho}^{4}r_{0}^{4} +\epsilon _{B}+\epsilon_{L}
\label{en:dens}
\end{eqnarray}
with $\varepsilon _{B}$ and $\varepsilon _{L}$ given by
\begin{equation}
\varepsilon_{B}=\sum_{B}\frac{1}{3\pi^2}\int_{0}^{k_{F,B}}k^2dk\sqrt{k^2+M_{eff,B}^2}
\end{equation}
\begin{equation}
\varepsilon_{L}=\sum_{l}\frac{1}{3\pi^2}\int_{0}^{k_{F,l}}k^2dk\sqrt{k^2+m_l^2}
\end{equation}
\begin{eqnarray}
P & = & \frac{1}{2}m_{\rho }
r_{0}^{2}+\frac{1}{2}m_{\omega}w_{0}^{2}+\frac{1}{4}c_3w_{0}^{4}-\frac{1}{2}m_{\delta }d^2+\frac{1}{2}m_{\phi }f_0^2+\\
\nonumber & + & \frac{1}{2}m_{\sigma^* }(s^{*})^2-U(s)
+\Lambda_{v}g_{\omega}^{2}g_{\rho}^{2}r_{0}^{2}\omega_{0}^{2 }+
\Lambda_{4}g_{\sigma}^{2}g_{\rho}^{2}r_{0}^{2}\varphi^{2} +
\frac{1}{24}\zeta g_{\rho}^{4}r_{0}^{4} +P_{B}+P_{L}
\label{ps:ress}
\end{eqnarray}
\begin{equation}
P_{B}=\sum_B\frac{1}{3\pi^2}\int_{0}^{k_{F,B}}\frac{k^4dk}{\sqrt{k^2+M_{eff,B}^2}}
\end{equation}
\begin{equation}
P_{L}=\sum_l\frac{1}{3\pi^2}\int_{0}^{k_{F,l}}\frac{k^4dk}{\sqrt{k^2+m_l^2}}
\end{equation}
%###################################################################
 \section{Neutron star parameters}
\begin{figure}
\fbox{\subfigure[]
{\includegraphics[width=7.5cm]{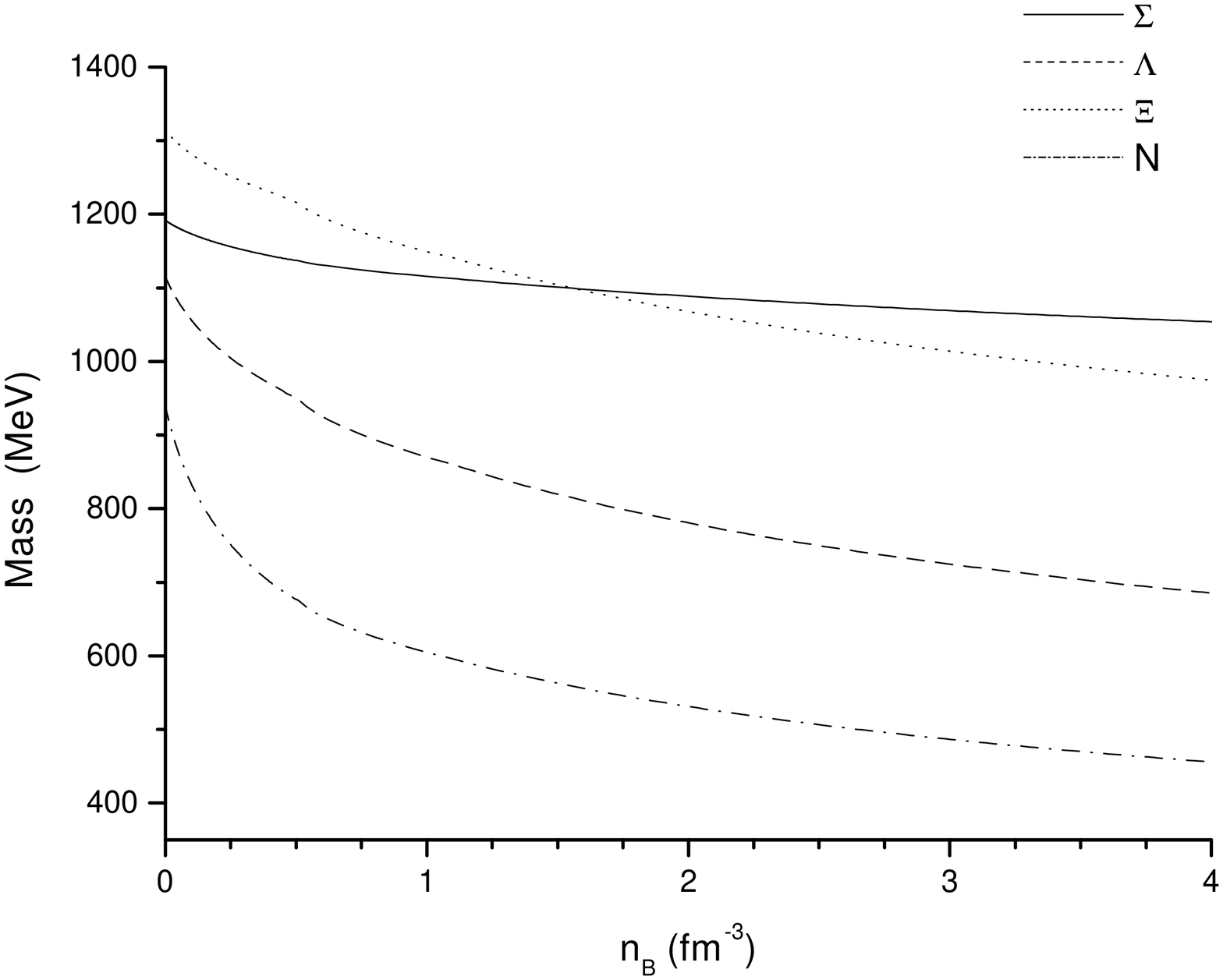}}}
\fbox{\subfigure[]
{\includegraphics[width=7.5cm]{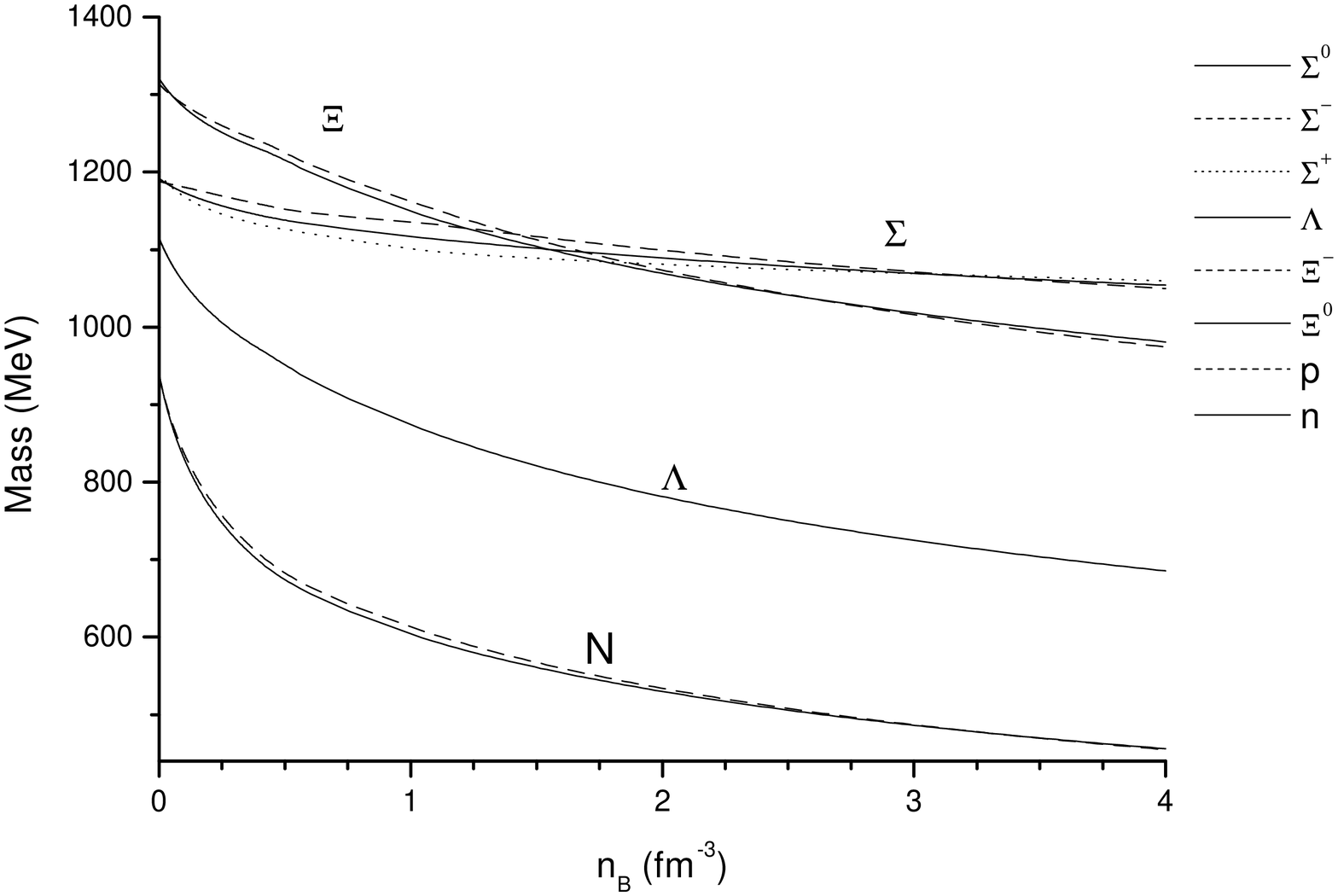}}}
\fbox{\subfigure[]
{\includegraphics[width=7.5cm]{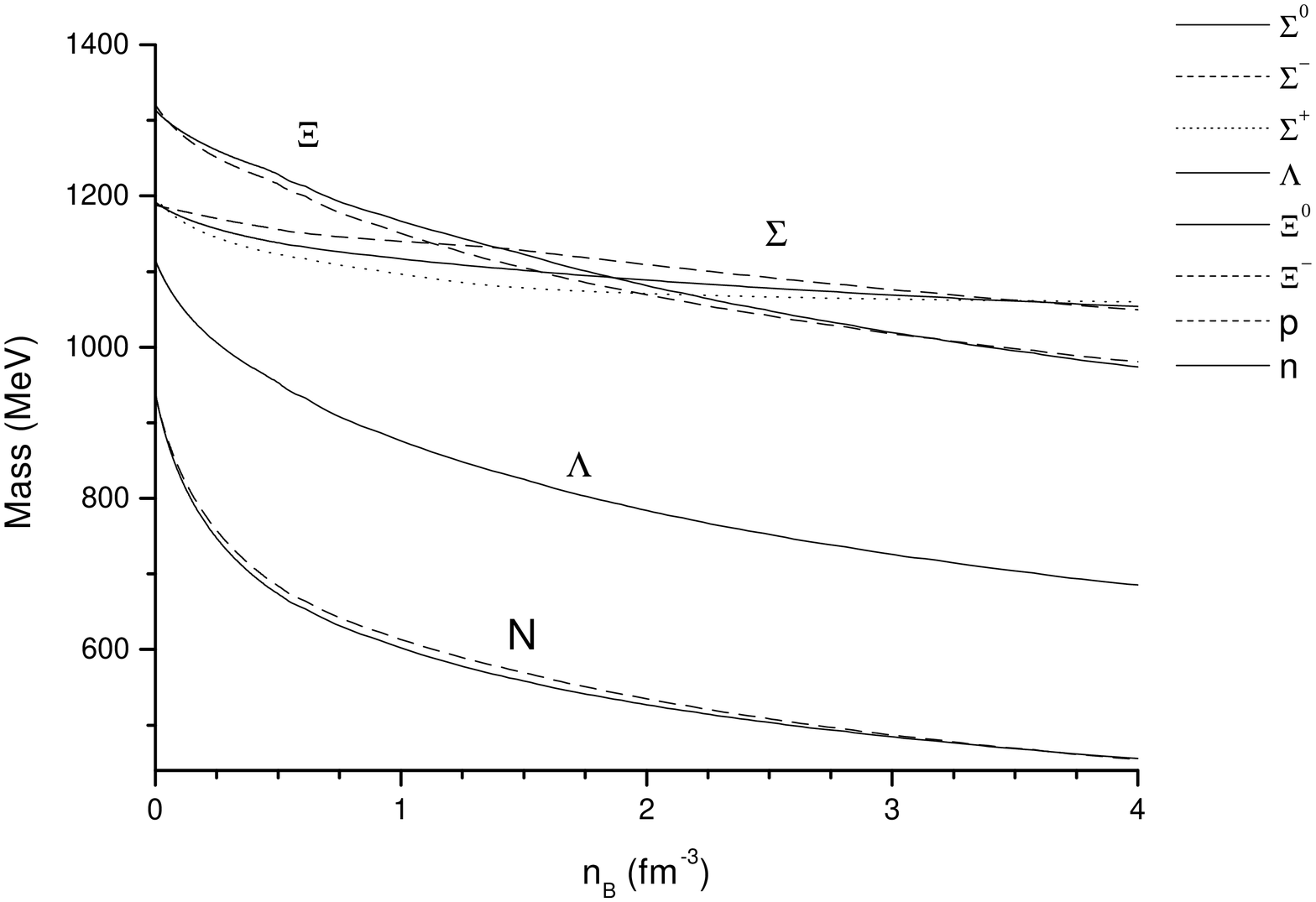}}}
\caption{Effective baryon masses as a function of  baryon number
density $n_B$. Fig.1a shows effective baryon masses obtained in
the model without $\delta $ meson (the parameter set I).In Fig.1b
and Fig.1c the effective baryon masses are splitted due to the
presence of $\delta$ meson. This effect is even more significant
in the case of more asymmetric matter (set III).}
 \label{fig:meff}
\end{figure}

Besides other properties of neutron stars the value of their
masses and radii are very sensitive to the chosen
model of strong interactions which in turn lead to the significant
constrains on the form of the equation of state of the neutron
star matter. In this paper three different groups of parameters
have been applied in order to examined neutron star properties. In
all of them the nucleon-hyperon $\Sigma$ interaction is assumed to
be repulsive. The first case marked as set I does not contain
$\delta$ meson, the second one (set II) do include $\delta$ meson,
in the third both $\delta$ meson and nonlinear vector meson
interactions are taken into account (set III). As the neutron star
matter is of sizable asymmetry the last case seems to be the most
adequate for the complete description of  the asymmetric neutron
star matter. The main effect of such an extension of the theory
becomes evident  studying  properties of the neutron star matter
especially baryon masses splitting and the form of the  equation
of state. Fig.\ref{fig:meff}  depicts the effective baryon masses
obtained for the three mentioned above cases as  functions of the
baryon number density $n_B$. For the second and third cases there
are noticeable in-medium baryon masses splitting for each
isomultiplet. This effect reaches the maximum value for the third
group of parameters, for the density span of $(3-4) \rho_0$ and
depends on the considered baryonic masses. Depending on the sign
of the third component of particular baryon isospin the $\delta$
meson interaction increases the proton and $\Xi^0$ effective
masses and decreases masses of neutron, $\Sigma^-$ and $\Xi^-$.
Contrary to this situation, for parameters marked as the set I,
the baryonic masses for the given isomultiplet remain degenerate
as it is shown in the left panel of Fig.\ref{fig:meff}(a). Having
obtained the effective baryonic masses  one can compare them with
those obtained in the Walecka  model. In the original Walecka
approach the nucleon effective mass rapidly diminishes its value
passing through zero and even become negative for higher
densities.
%%%%%%%%%%%%%%%%%%%%%%%%%%%%%%%%%%%%%%%%%%%%%%%%%%%%%%%%%%%
%%%%%%%%%%%%%%%%%%%%%%%%%%%%%%%%%%%%%%%%%%%%%%%%%%%%%%%%%%
Throughout the effective baryonic masses the asymmetry of the system alters
baryon chemical potentials what realizes
in characteristic modification of the appearance, abundance and distributions of the individual flavors.
This is evident comparing the results obtained for the three  mentioned above parameter sets.
%%%%%%%%%%%%%%%%%%%%%%%%%%%%%%%%%%%%%%%%%%%%%%%%%%%%%
\begin{figure}
\fbox{\subfigure[]
{\includegraphics[width=7.5cm]{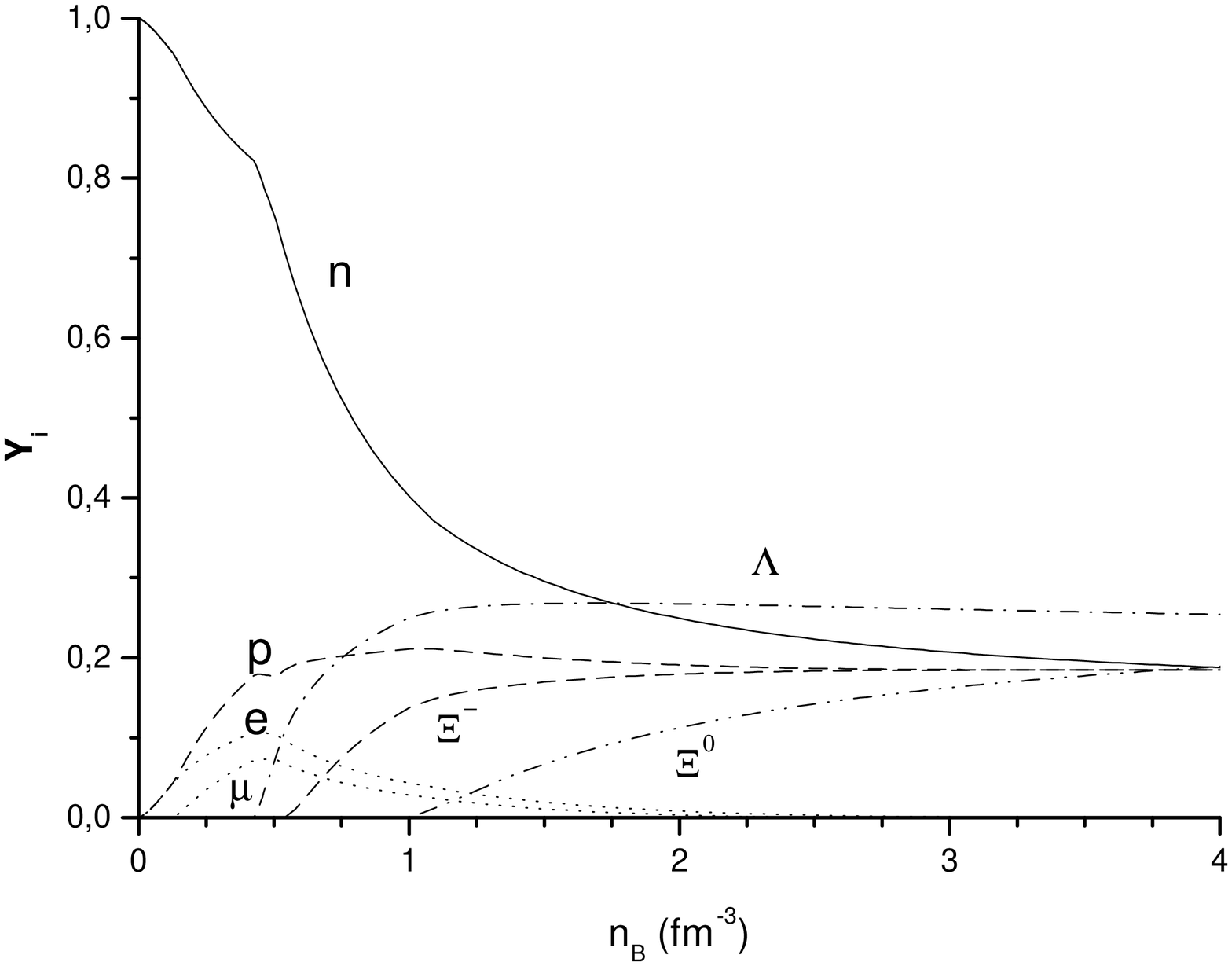}}}
\fbox{\subfigure[]
{\includegraphics[width=7.5cm]{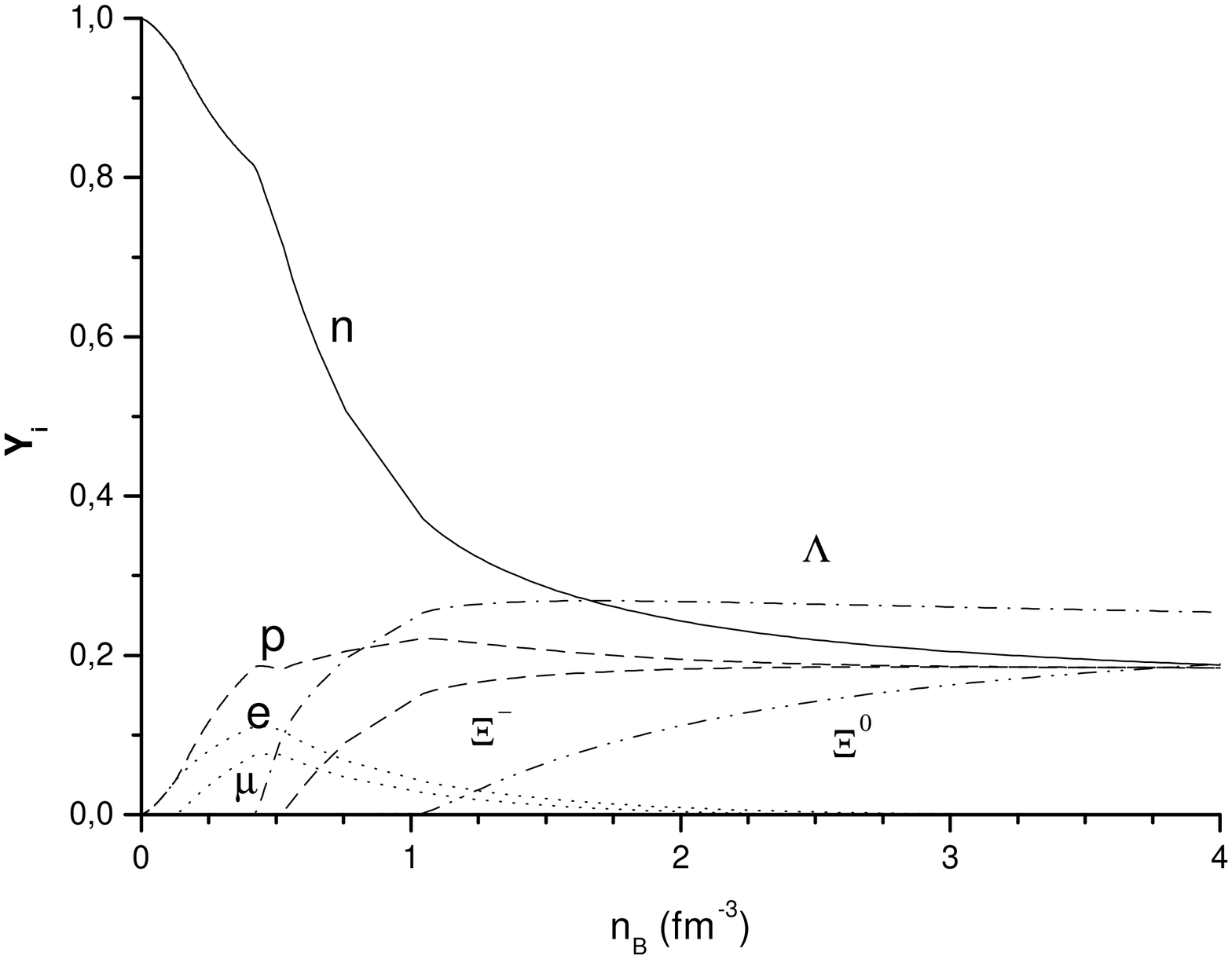}}}
\fbox{\subfigure[]
{\includegraphics[width=7.5cm]{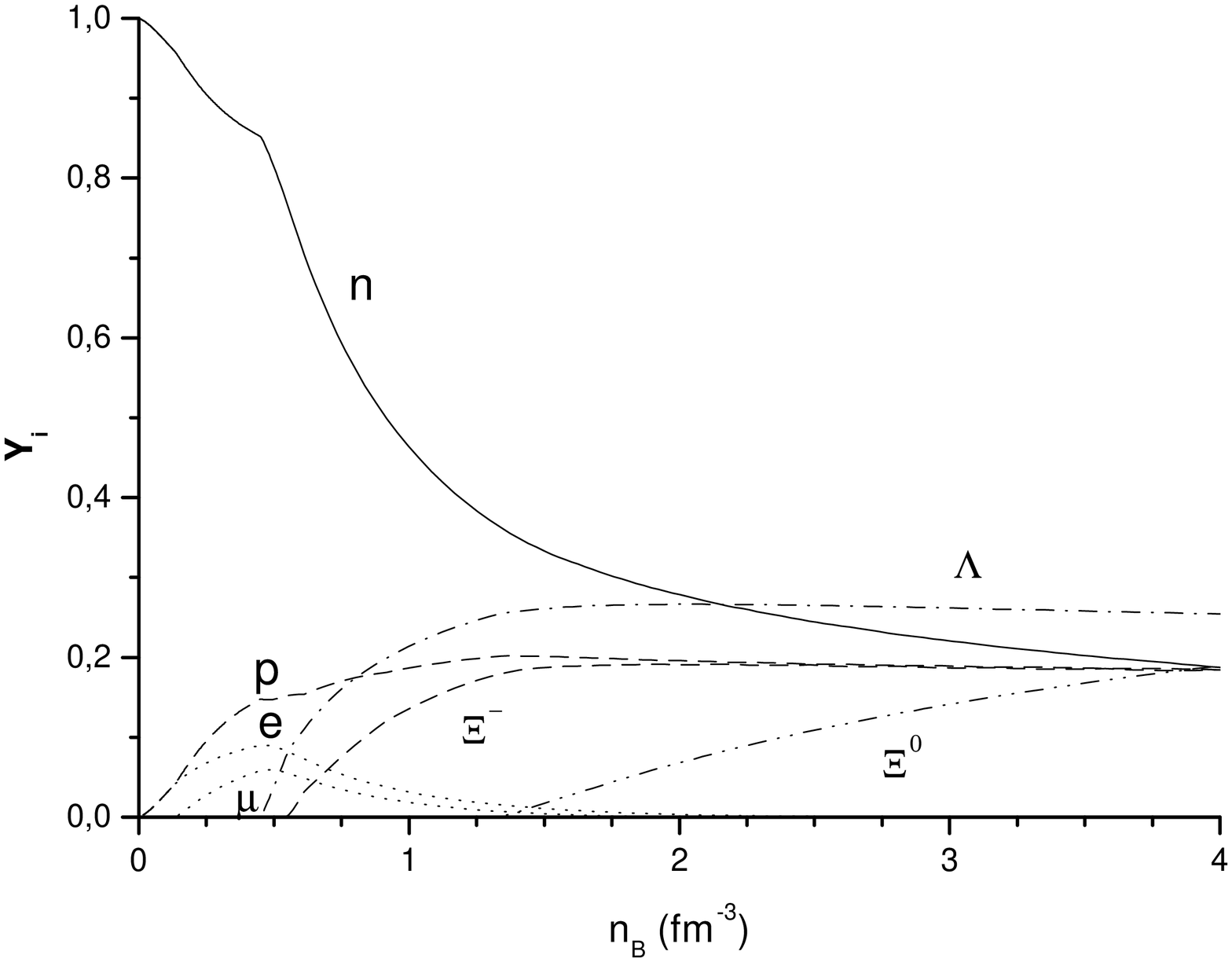}}}
\caption{The equilibrium compositions as functions of baryon
number density $n_{B}$ for given parameter sets. Fig.2a shows
results obtained for the parameter set I. Fig.2a and 2b depict
equilibrium compositions of matter which contain $\delta$ meson
(set II) and additional vector meson interactions (set III).}
\label{fig:partn}
\end{figure}
\begin{figure}\fbox{
\includegraphics[width=15cm]{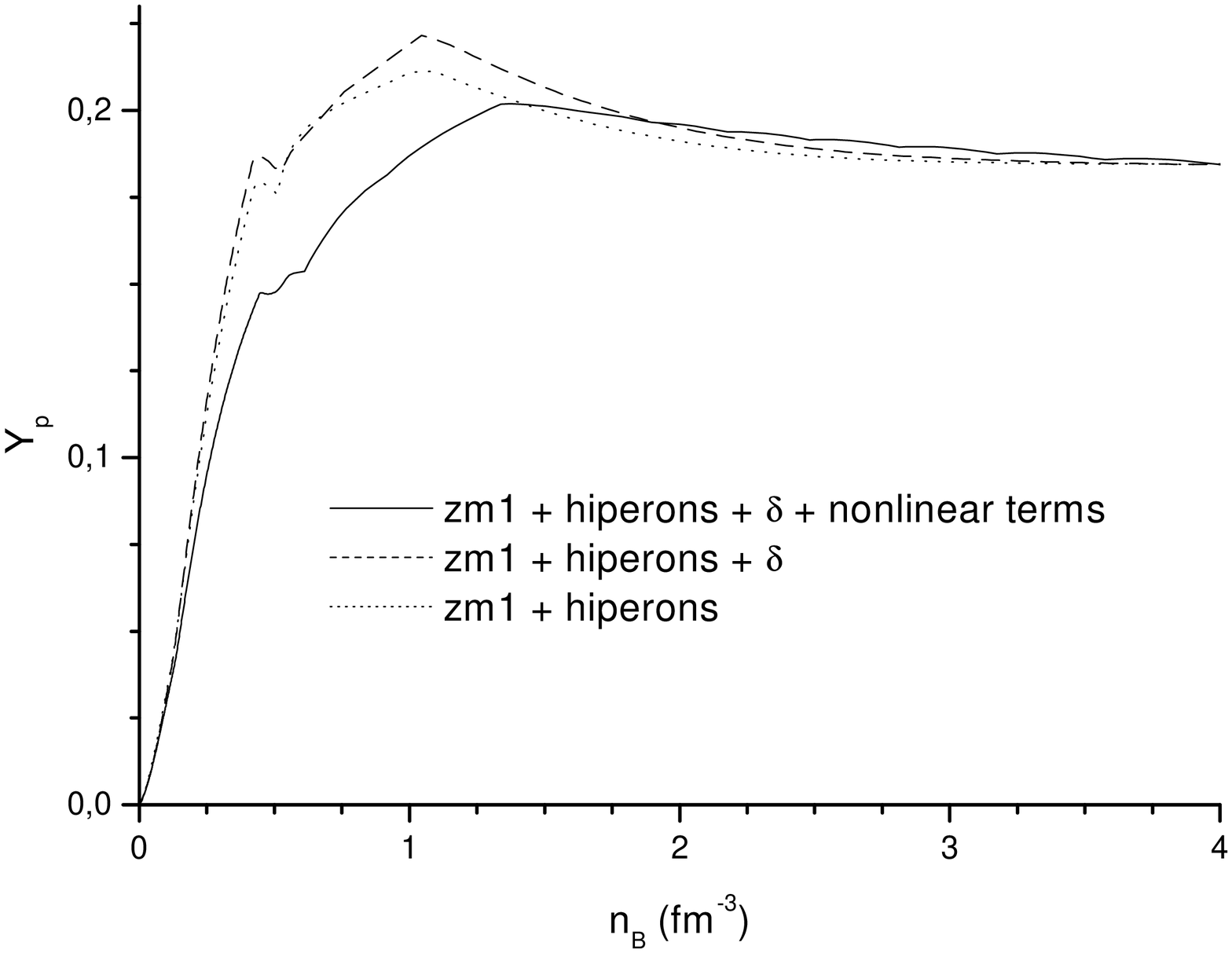}}
\caption{Proton fractions $Y_p$, as functions of  baryon number
density $n_{B}$. Solid line represents the proton fraction for the
most asymmetric matter. Dotted and dashed lines are obtained for
parameter set I and II respectively.} \label{fig:proton}
\end{figure}
%%%%%%%%%%%%%%%%%%%%%%%%%%%%%%%%%%%%%%%%%%%%%%%%%%%%%%%%
Fig.\ref{fig:partn} presents fractions of particular baryon
species $Y_B$ as  functions of baryon number density $n_B$.
Starting the analysis of these graphs from low and moderate
densities it is evident that at very low density neutrons are the
most abundant baryon and a star resembles the pure neutron one.
For higher densities protons and electrons and then muons emerge.
Fig.2 points that the first strange baryon that appears at $\rho
=2.5 \times \rho_0$ is the $\Lambda$, it is followed by $\Xi^-$
and $\Xi^0$. In the presence of $\delta$ meson and in the case
when nonlinear vector meson interactions are included the sequence
of appearance of hyperons is the same as in the first case,
however shifted towards higher densities. Due to the repulsive
potential of $\Sigma$ hyperons their onset points are possible at
very high densities which are not relevant for neutron stars.
Taking into account results obtained when the attractive
nucleon-hyperon $\Sigma$ potential is assumed $\Sigma^0$ appear as
the first strange baryon and than $\Lambda$ and $\Sigma^+$.
$\Xi^0$ emerge at the density above $7\rho_0$, whereas other
hyperons at even higher densities.\\ The appearance of charged
hyperons permits the lower lepton contents and thus charge
neutrality tends to be guaranteed without lepton contribution.
This  kind of deleptonization in the case of the set II and III
parameterizations similarly to the baryon distributions takes
place at higher densities. Larger effective baryon masses cause
the shift of the given hyperon onset point especially for the
charged ones towards  higher densities. The emergence of $\Xi^-$
hyperons through the condition of charge neutrality affects the
lepton fraction and causes a drop in their contents. The proton
fraction $Y_p$ has a crucial role in neutron stars cooling
history. At a certain critical value of $Y_p$ the direct URCA
processes for neutrino emissions are allowed. The condition which
has to be satisfied for the URCA process is given by the relation
between Fermi momenta of nucleons and electrons
\begin{equation}
k_{Fp}+k_{Fe}\geq k_{Fn}
\end{equation}
The threshold proton fraction for this process is about 0.11 in
the case when the neutron star matter consists only of nucleons
and electrons. Inclusion of muons increases $Y_p$ to the value
$\sim 0.14$. The proton fractions which are obtained for the
presented parameter groups are shown in Fig.\ref{fig:proton}. The
third parameter set gives the lower value of $Y_p$. \clearpage
Baryon distributions are strictly connected with the behavior of
meson fields. The values of scalar and vector meson fields
influence the onset points of individual hyperon species. Fig.
\ref{fig:fieldsn} shows meson fields as functions of the baryon
number density.\\
%%%%%%%%%%%%%%%%%%%%%%%%%%%%%%%%%%%%%%%%%%%%%%%%%%%%%%%%%%
\begin{figure}
\fbox{\subfigure[]
{\includegraphics[width=7.5cm]{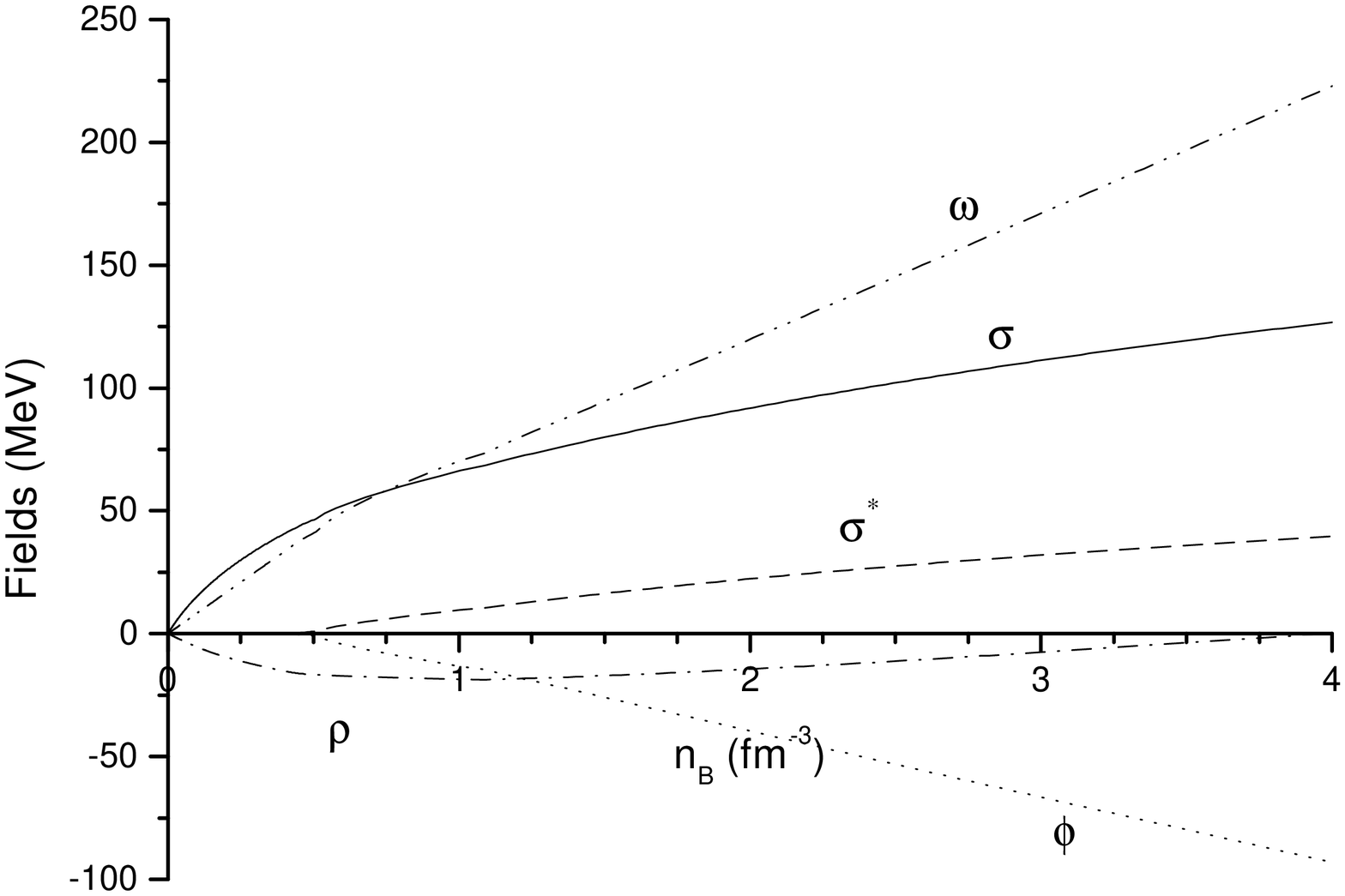}}}
\fbox{\subfigure[]
{\includegraphics[width=7.5cm]{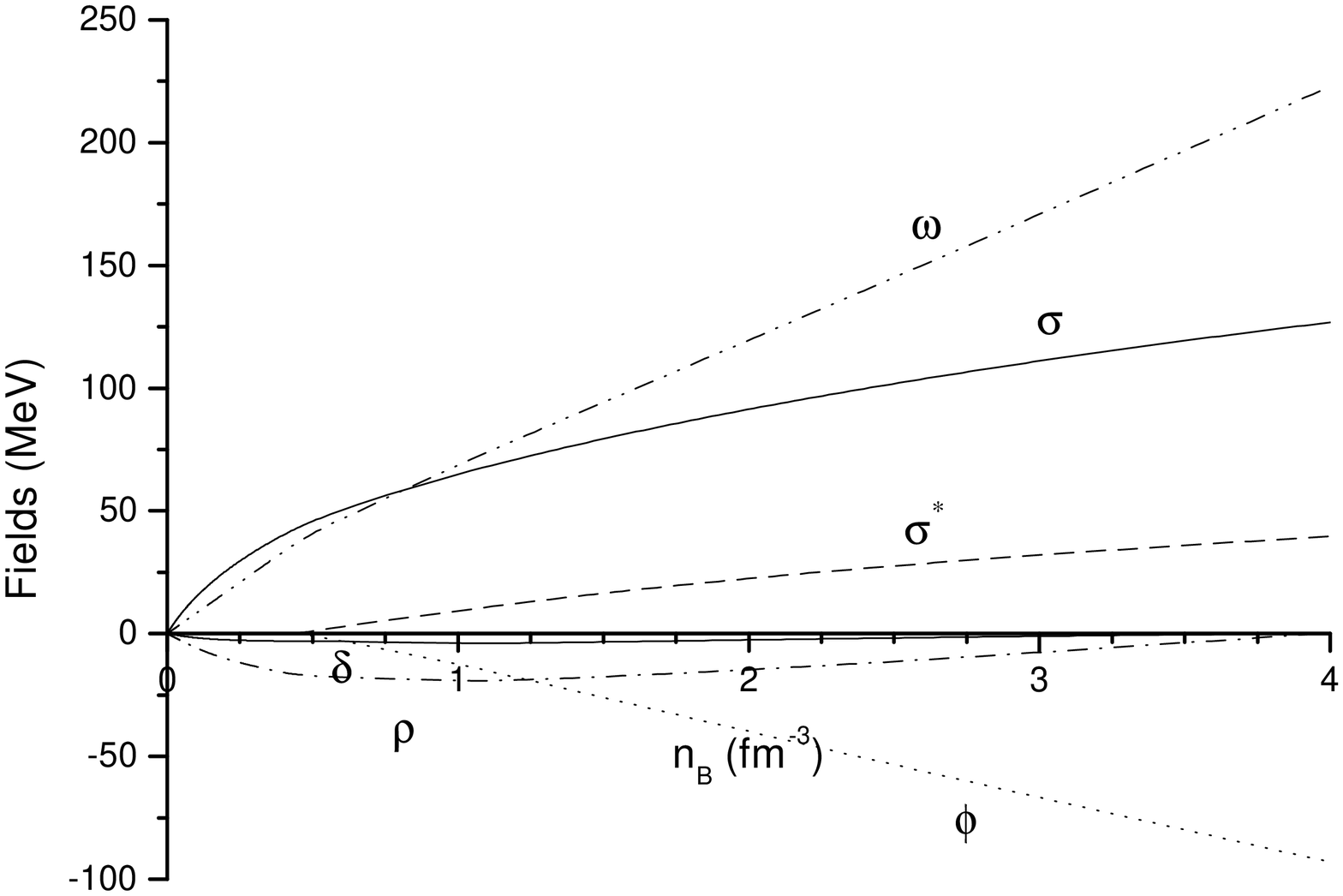}}}
\fbox{\subfigure[]
{\includegraphics[width=7.5cm]{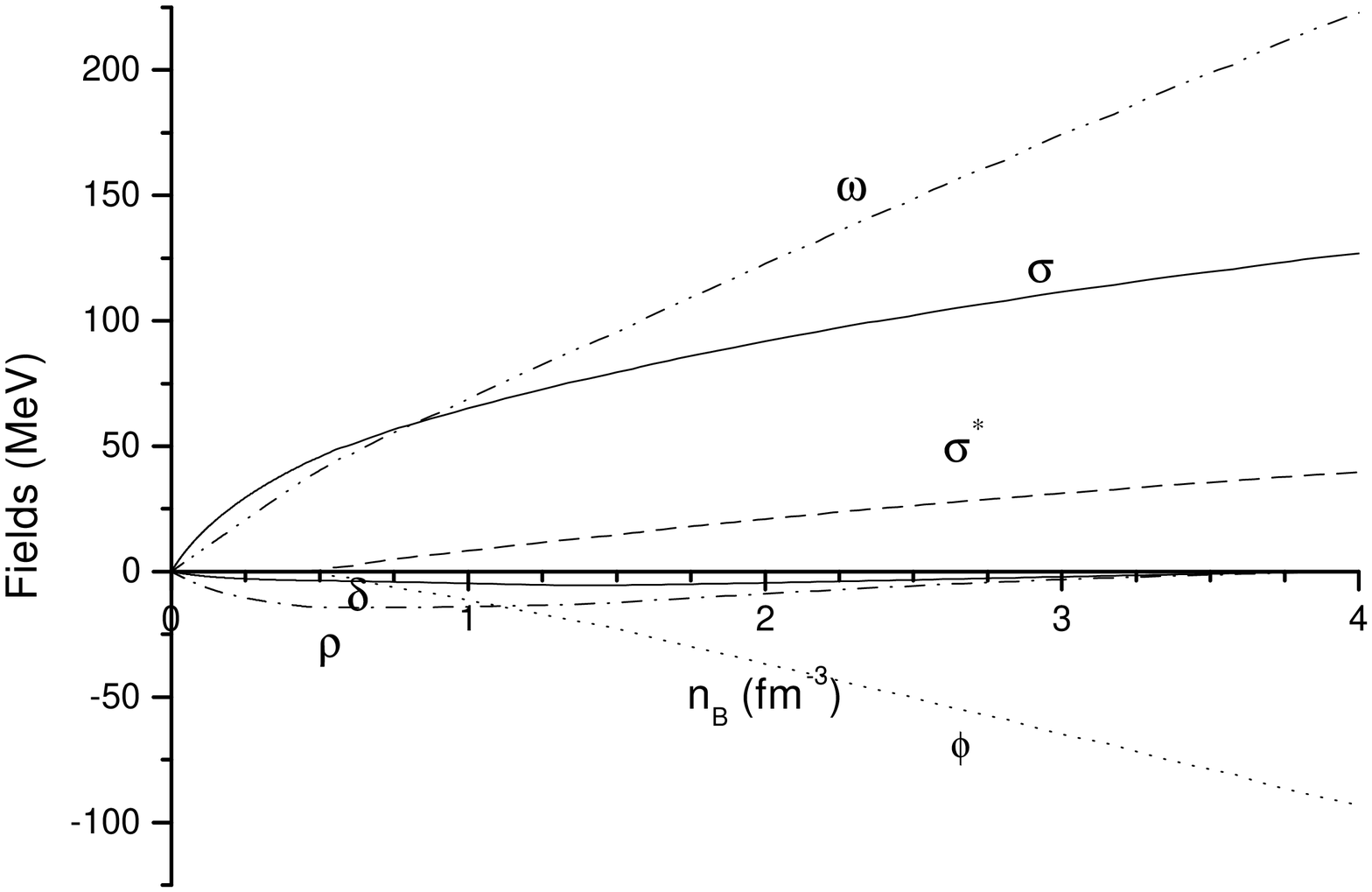}}}
\caption{The meson fields as functions of baryon number density
$n_B$ for given parameter sets.} \label{fig:fieldsn}
\end{figure}
%%%%%%%%%%%%%%%%%%%%%%%%%%%%%%%%%%%%%%%%%%%%%%%%%%%%%%%%%
\begin{figure}\fbox{
\includegraphics[width=15cm]{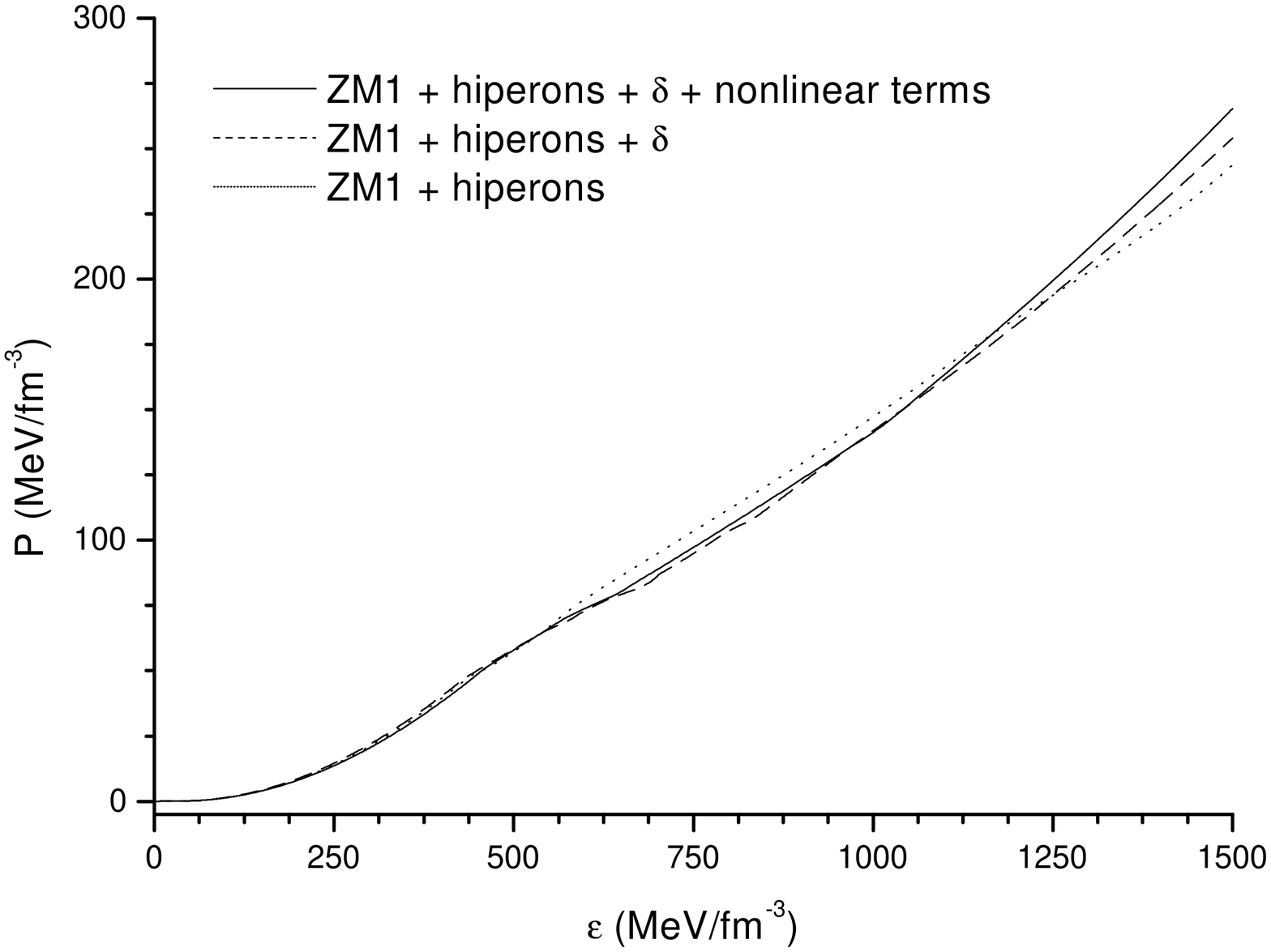}}
\caption{The equation of state (EOS). Straight, dashed and dotted
lines represent models constructed with the use of parameter sets
I,II and III respectively.} \label{fig:eos}
\end{figure}
%%%%%%%%%%%%%%%%%%%%%%%%%%%%%%%%%%%%%%%%%%%%%%%%%%%%%%%%%
The calculated form of the equation of state (EOS) determines the physical state and
composition of matter at high densities and is presented in Fig.\ref{fig:eos}. The
relative hadron-lepton composition calculated for all parameter
groups can be  analyzed through the density dependence of the
asymmetry parameter $f_a$ and the parameter $f_s$ which is connected
with the strangeness contents. Fig.\ref{fig:asym} presents
both parameters as  functions of the baryon number density $n_B$.
As the density increases the asymmetry parameter decreases. The
third parameter group gives the higher value of the asymmetry
parameter.  For parameter sets I and II the asymmetry is
comparable. The strangeness contents increases with the density.
The inclusion of $\delta$ meson and nonlinear vector meson
interactions results in the lowest hyperon contents.\\
%%%%%%%%%%%%%%%%%%%%%%%%%%%%%%%%%%%%%%%%%%%%%%%%%%%%%%%%%%%%%%
\begin{figure}\fbox{
\includegraphics[  width=15cm]{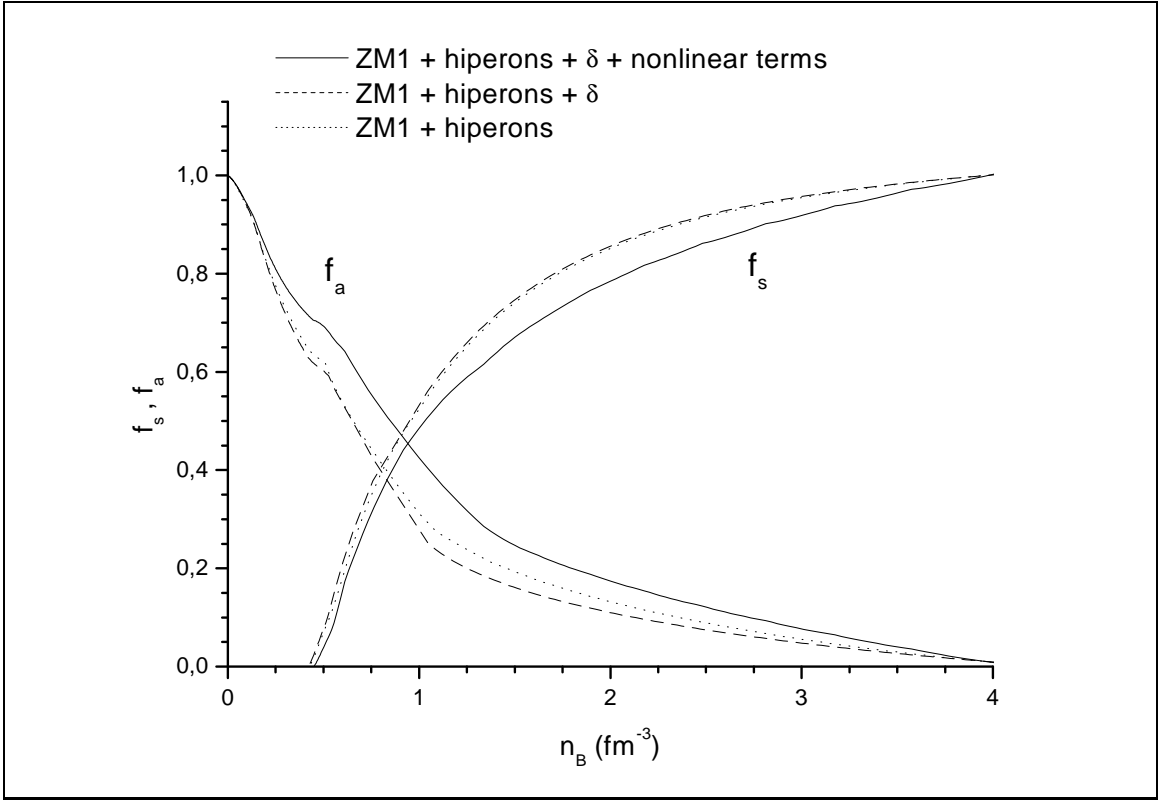}}
\caption{The asymmetry parameter $f_{a}$ and the strangeness
parameter $f_{s}$  as functions of baryon number density $n_{B}$.}
\label{fig:asym}
\end{figure}
%%%%%%%%%%%%%%%%%%%%%%%%%%%%%%%%%%%%%%%%%%%%%%%%%%%%%%%%%%%%%%%%%%
The obtained  form of the equation of state serves as an input to
the Oppenheimer-Volkoff equations and  determines the structure of
spherically symmetric stars.
\begin{eqnarray}
\frac{dP(r)}{dr}=-\frac{Gm(r)\rho(r)}{r^2}\frac{(1+\frac{P(r)}{\rho(r)})(1+\frac{4\pi r^3P(r)}{m(r)})}{1-\frac{2Gm(r)}{r}}, \\ \nonumber
\frac{dm(r)}{dr}=4\pi r^2\rho (r).
\end{eqnarray}
Numerical solutions of these equations allow to construct the
mass-radius relations. These relations for the chosen form of the
equations of state are presented in Fig.\ref{fig:maspr}. Arrows
represent points at which hyperons emerge whereas asterisks the
configurations for which URCA processes start to proceed. For the
parameter set III these two marks are in the same position. This
figure also shows that  $\delta$ meson itself and the
additional nonlinear vector meson interactions change the maximal
mass but not in a significant way. However, there are visible
differences in the maximal radius configurations. For comparison,
the mass-radius relation for a quark star is enclosed.
%%%%%%%%%%%%%%%%%%%%%%%%%%%%%%%%%%%%%%%%%%%%%%%%%%%%%%%%%%%%%%%%%%
\begin{figure}\fbox{
\includegraphics[width=15cm]{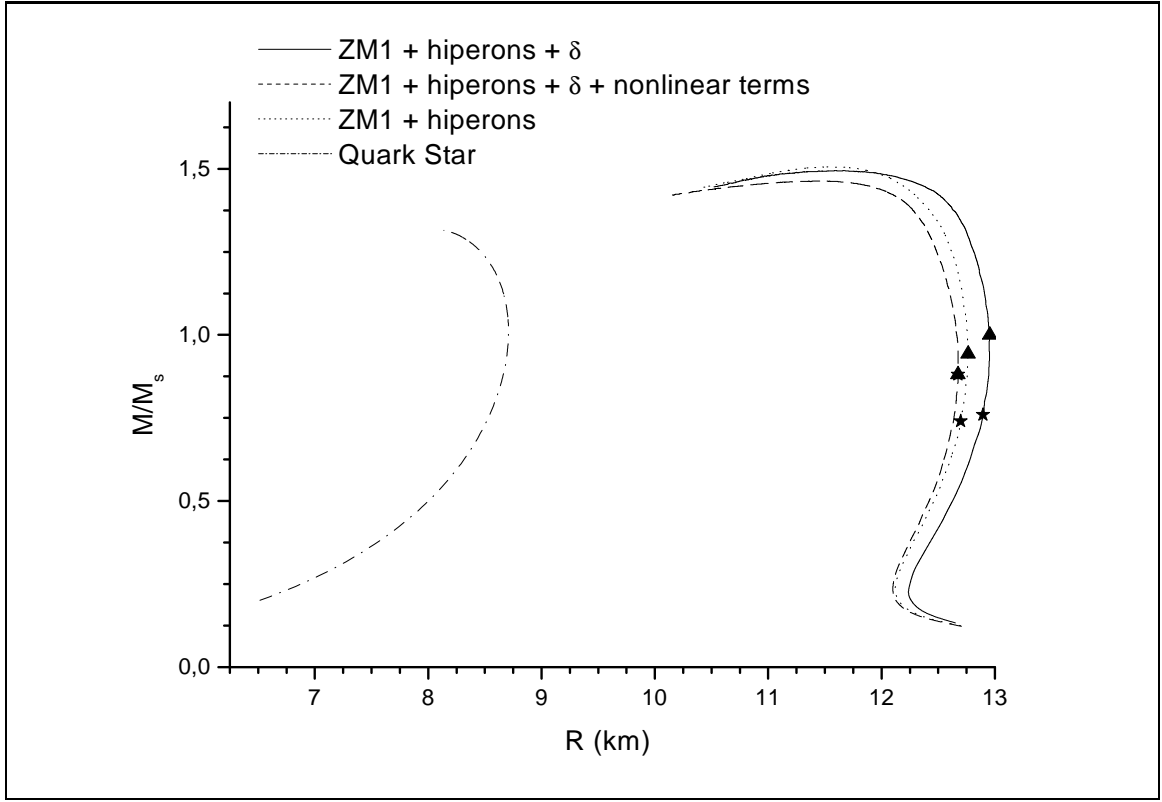}}
\caption{Mass radius relations. Asterisks represent points where
direct URCA processes start whereas arrows the points where
hiperons emerge.} \label{fig:maspr}
\end{figure}
%%%%%%%%%%%%%%%%%%%%%%%%%%%%%%%%%%%%%%%%%%%%%%%%%%%%%%%%%%%%%%%%%
In Fig.\ref{fig:masro} the obtained neutron star masses as functions of central densities are presented.
%%%%%%%%%%%%%%%%%%%%%%%%%%%%%%%%%%%%%%%%%%%%%%%%%%%%%%%%%%%%%%%%%
\begin{figure}\fbox{
\includegraphics[width=15cm]{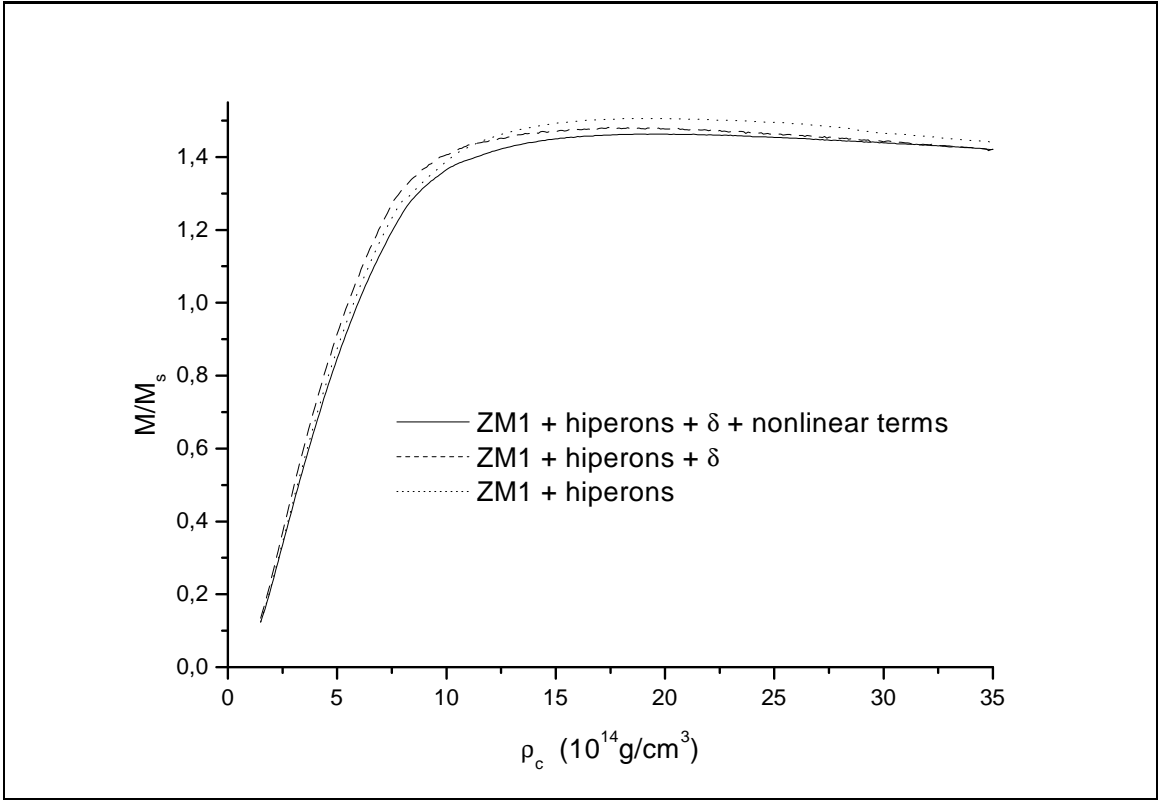}}
\caption{The neutron star masses as functions of the central
density of neutron stars.} \label{fig:masro}
\end{figure}
%%%%%%%%%%%%%%%%%%%%%%%%%%%%%%%%%%%%%%%%%%%%%%%%%%%%%%%%%%%%%%%%%%%%
\clearpage White dwarfs and neutron stars are purely
gravitationally bound compact stars. The gravitational binding
energy of a relativistic star is defined as a difference between
its gravitational and baryon masses.
\begin{equation}
E_{b,g}=(M_{p}-m(R))c^{2}\end{equation} where \begin{equation}
M_{p}=4\pi \int
_{0}^{R}drr^{2}(1-\frac{2Gm(r)}{c^{2}r})^{-\frac{1}{2}}\rho
(r)\end{equation} Of considerable relevance is the numerical
solution of the above equation for the selected EOS. Results are
shown in Fig. \ref{fig:engr}.
%%%%%%%%%%%%%%%%%%%%%%%%%%%%%%%%%%%%%%%%%%%%%%%%%%%%%%%%%%%%%%%%%%%%
\begin{figure}\fbox{
\includegraphics[width=15cm]{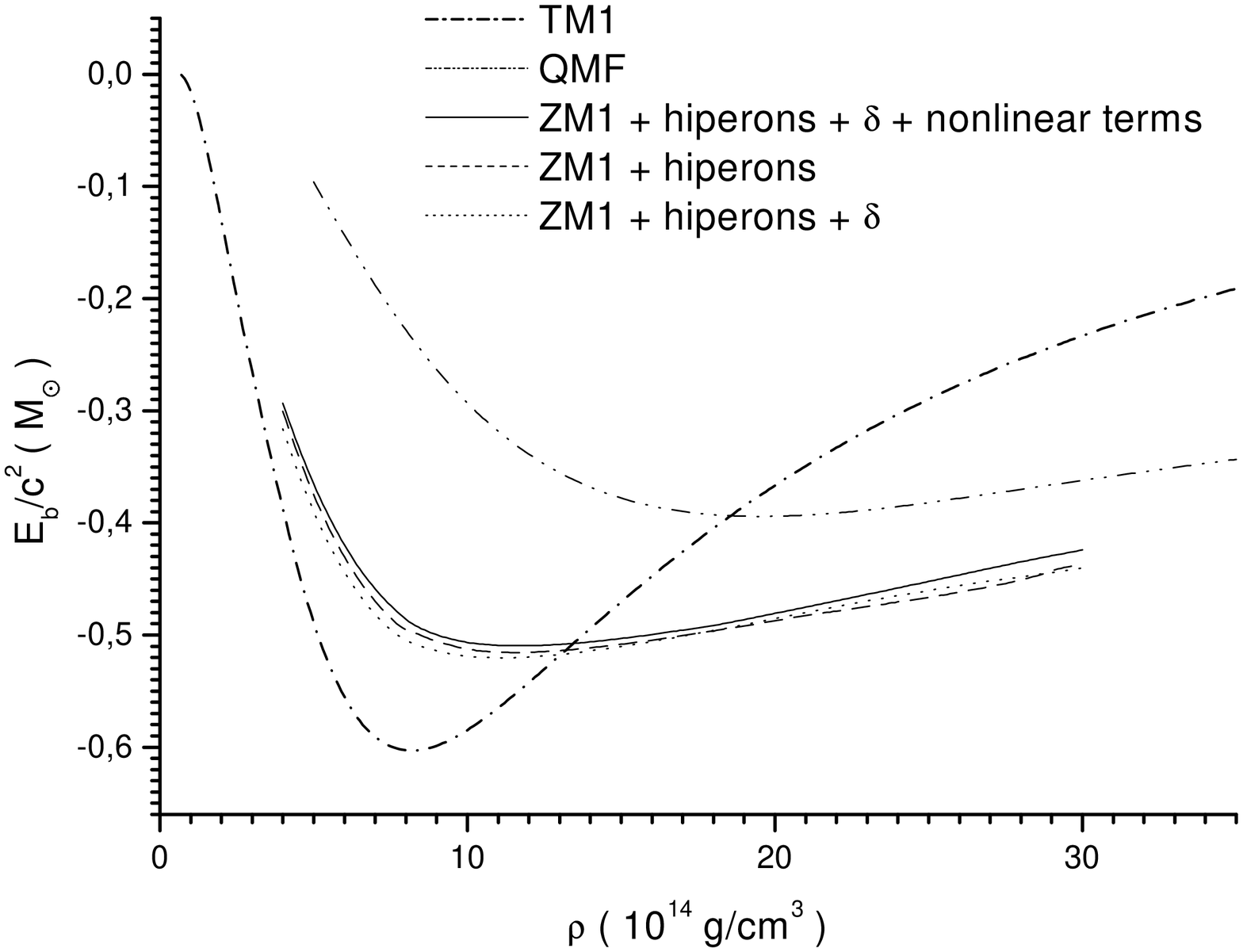}}
\caption{Gravitational binding energy as a function of central
density of the neutron star.} \label{fig:engr}
\end{figure}
%%%%%%%%%%%%%%%%%%%%%%%%%%%%%%%%%%%%%%%%%%%%%%%%%%%%%%%%%%%%%%%%%%%%
For the sake of completeness calculations have been done not only
for the presented parameter sets but also for the original TM1
parameter set and for a quark star. From this figure it is evident
that at moderate densities the standard neutron star model
calculated with the use of the TM1 parameter set is the most
energetically favorable configuration. At higher densities this
standard RMF approach become unsatisfactory. There is room for the
improvement of this theory by the introduction of other degrees of
freedom. The comparison of the models which  contain hyperons with
the Walecka one is performed. At high densities the models with
hyperons are more favorable than the model calculated with the use
of the TM1 parameter set. This is directly connected with the value of
effective baryon masses, obtained with the use of the
TM1 parameters, which diminish considerably their values
which after passing through zero  can take even the negative
value. This can be interpreted as a break down of the theory and
has its confirmation in the form of the gravitational binding
energy. Fig.\ref{fig:engr}
 depicts the gravitational binding energies as  functions
of the density. From this figure it is evident that the  models
without hyperons are energetically favorable at the density range
relevant for neutron stars. The situation changes for higher
densities and now models with hyperons become energetically
favorable. There are theoretical suggestions about the existence
of a quark  star but Fig.\ref{fig:engr} exhibits that a quark
star configuration should appear at very
high densities. \\
The gained solutions of the structure equations allows as to carry
out a similar analysis of the onset point, abundance and
distributions of the individual hadron and lepton species  but now
as  functions of the star radius $R$. Comparing results obtained
for the three presented above cases (parameter set I, II and III)
one can come to the conclusion that in all the cases the
assumption of the repulsive $\Sigma$ interaction shifts the onset
point of $\Sigma$ hyperons to very high densities and they do not
appear in neutron star interiors calculated in this model. Two
characteristic configurations have been considered. Namely the one
connected with the maximum mass configuration and the second with
that of the maximum radius. The very compact hyperon core which
emerges in the interior of the maximum mass configuration consist
of $\Xi^0$, $\Xi^-$ and $\Lambda$ hyperons only for the second parameter
set. The hyperon population is reduced to  $\Lambda$ and $\Xi^-$
for the parameter sets I and III. This can be seen in
Fig.\ref{fig:partm} and confirm by the
behavior of meson fields in the neutron star. Especially by the
hidden strange $\sigma^*$ and $\phi$ mesons which appear in the
vicinity of the neutron star center (Fig.\ref{fig:fieldsm} and
Fig.\ref{fig:fieldsr}). For the maximum radius configuration
for all groups of parameters  hyperons do not emerge in the interior of
neutron stars.
%%%%%%%%%%%%%%%%%%%%%%%%%%%%%%%%%%%%%%%%%%%%%%%%%%%%%%%%%%%%%%
\begin{figure}
\fbox{\subfigure[]
{\includegraphics[width=7.5cm]{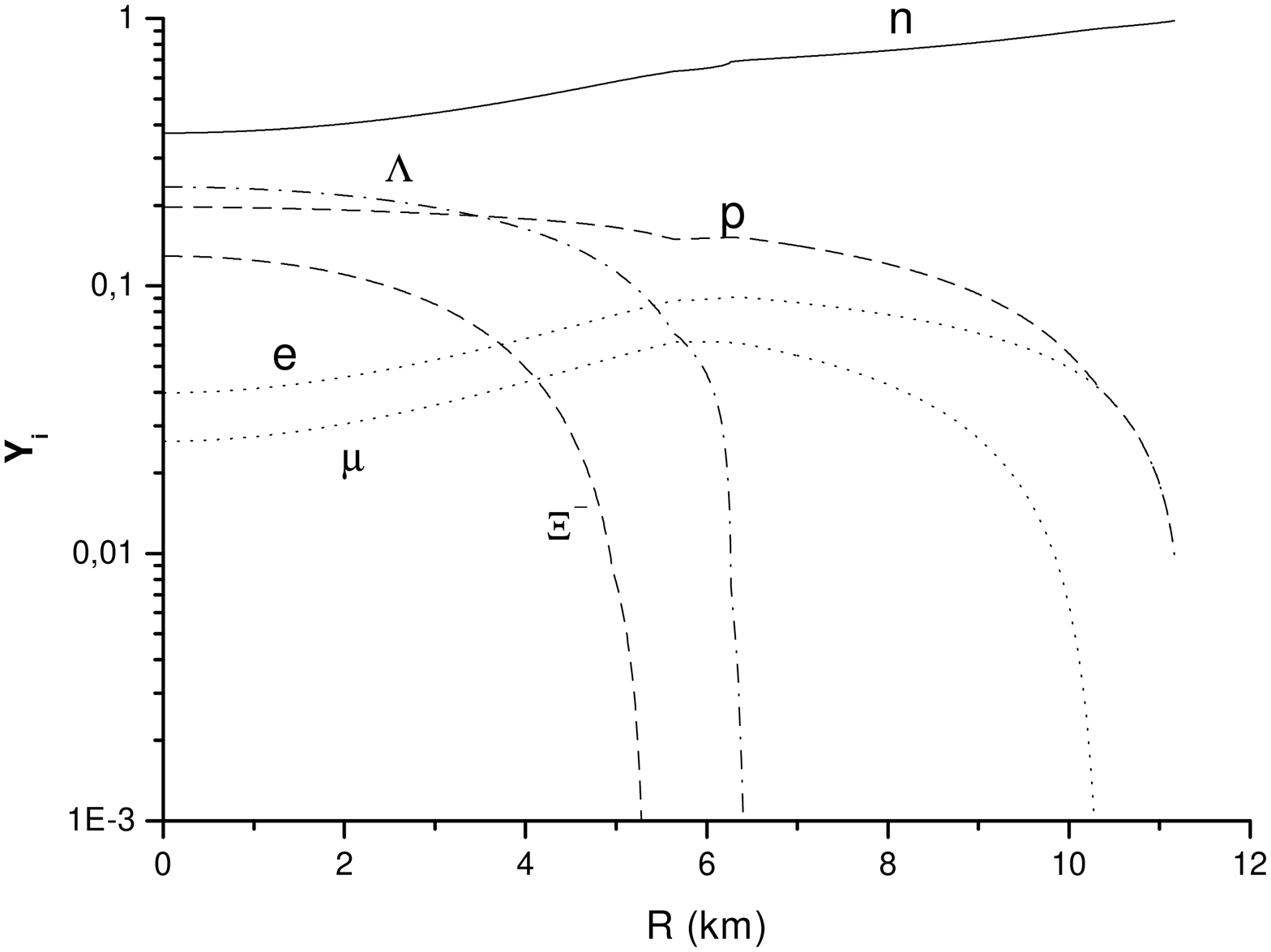}}}
\fbox{\subfigure[]
{\includegraphics[width=7.5cm]{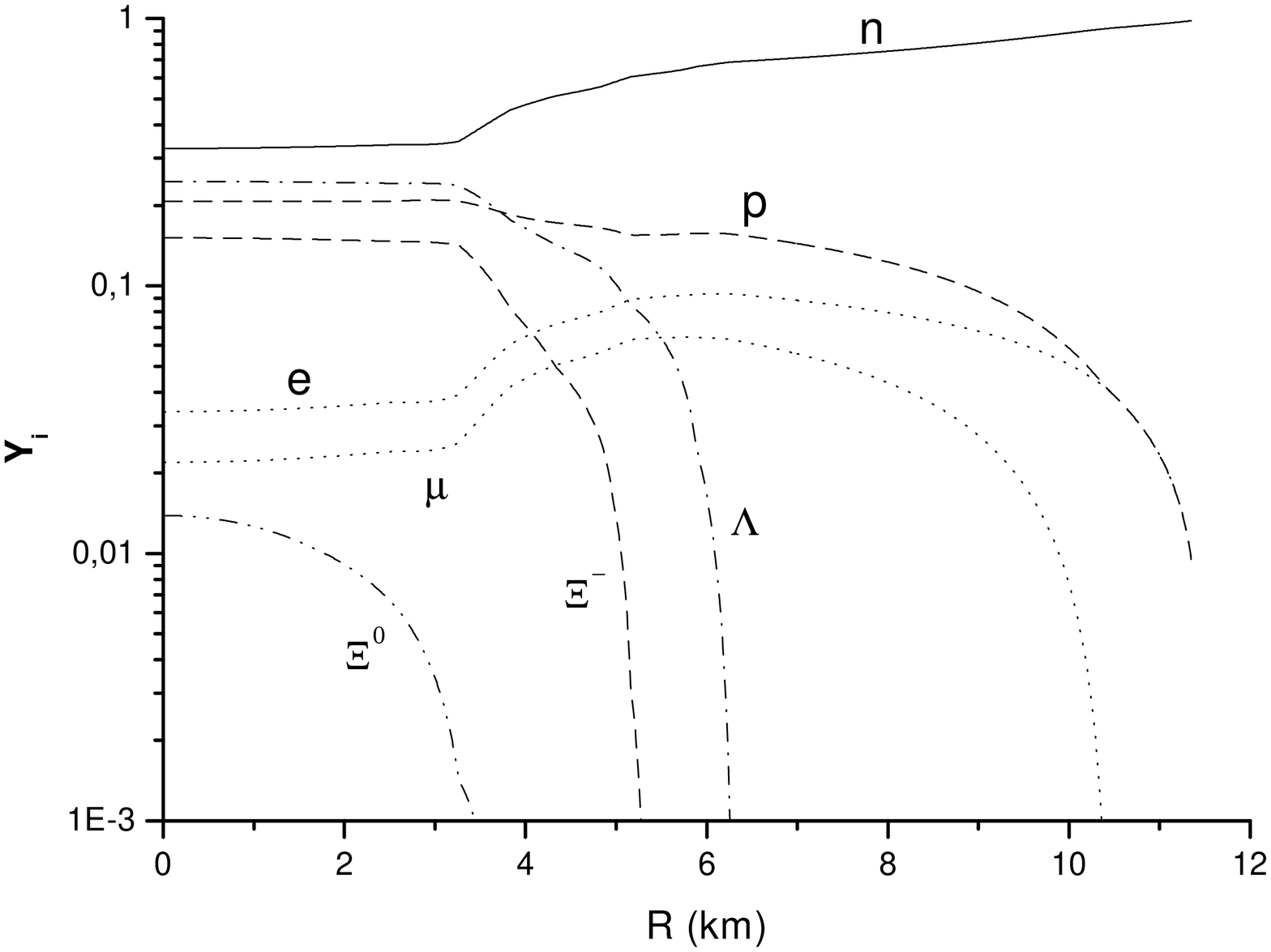}}}
\fbox{\subfigure[]
{\includegraphics[width=7.5cm]{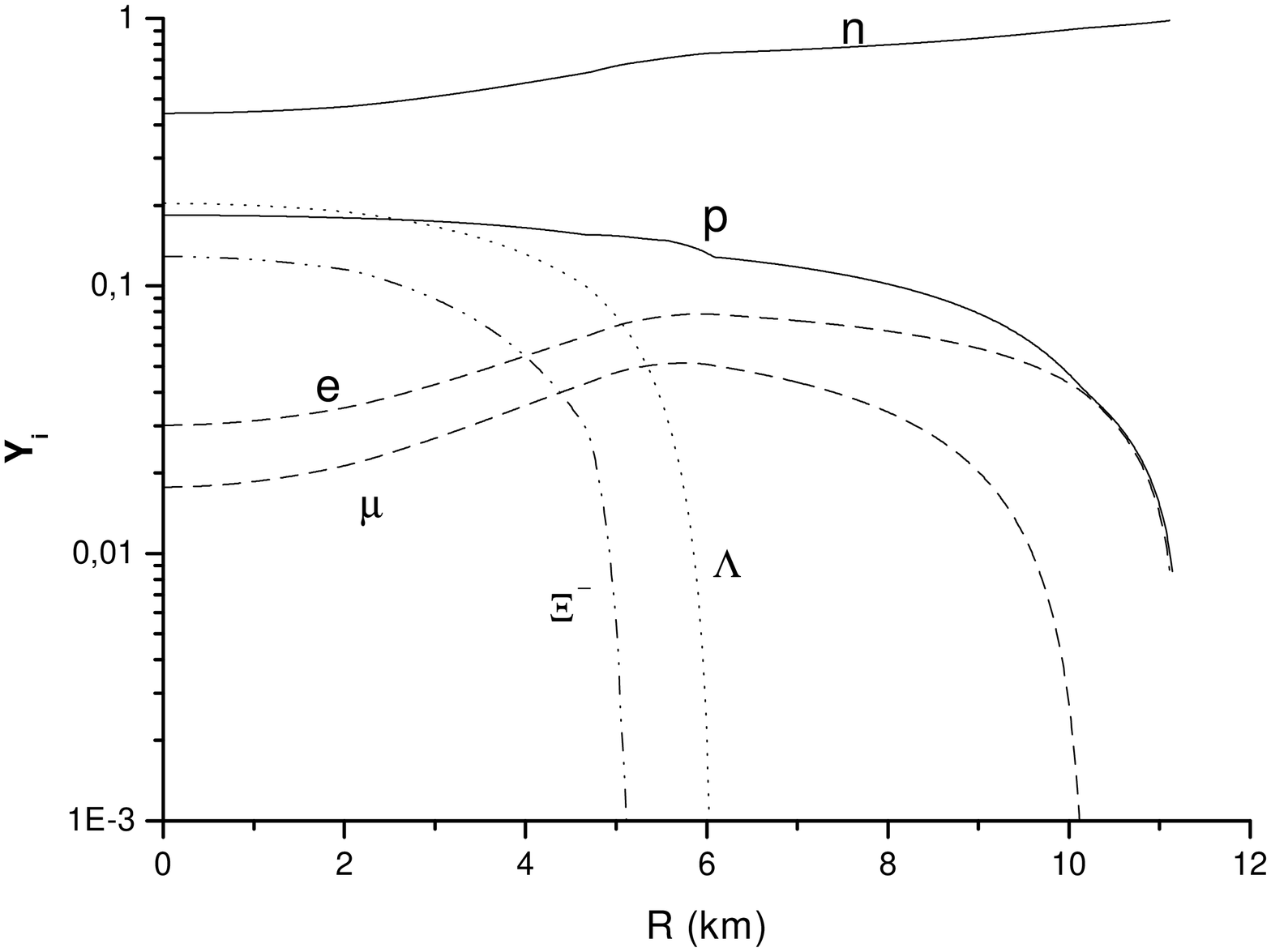}}}
\caption{The equilibrium compositions  for the maximum mass
configuration as a function of the star radius. Fig.10a is
constructed for the parameter set I, Fig.10b for the parameter set
II. Panel (c) represents results obtained for the parameter set
III.} \label{fig:partm}
\end{figure}
%%%%%%%%%%%%%%%%%%%%%%%%%%%%%%%%%%%%%%%%%%%%%%%%%%%%%%%%%%%%%
%%%%%%%%%%%%%%%%%%%%%%%%%%%%%%%%%%%%%%%%%%%%%%%%%%%%%%%%%%%%%
\begin{figure}
\fbox{\subfigure[]
{\includegraphics[width=7.5cm]{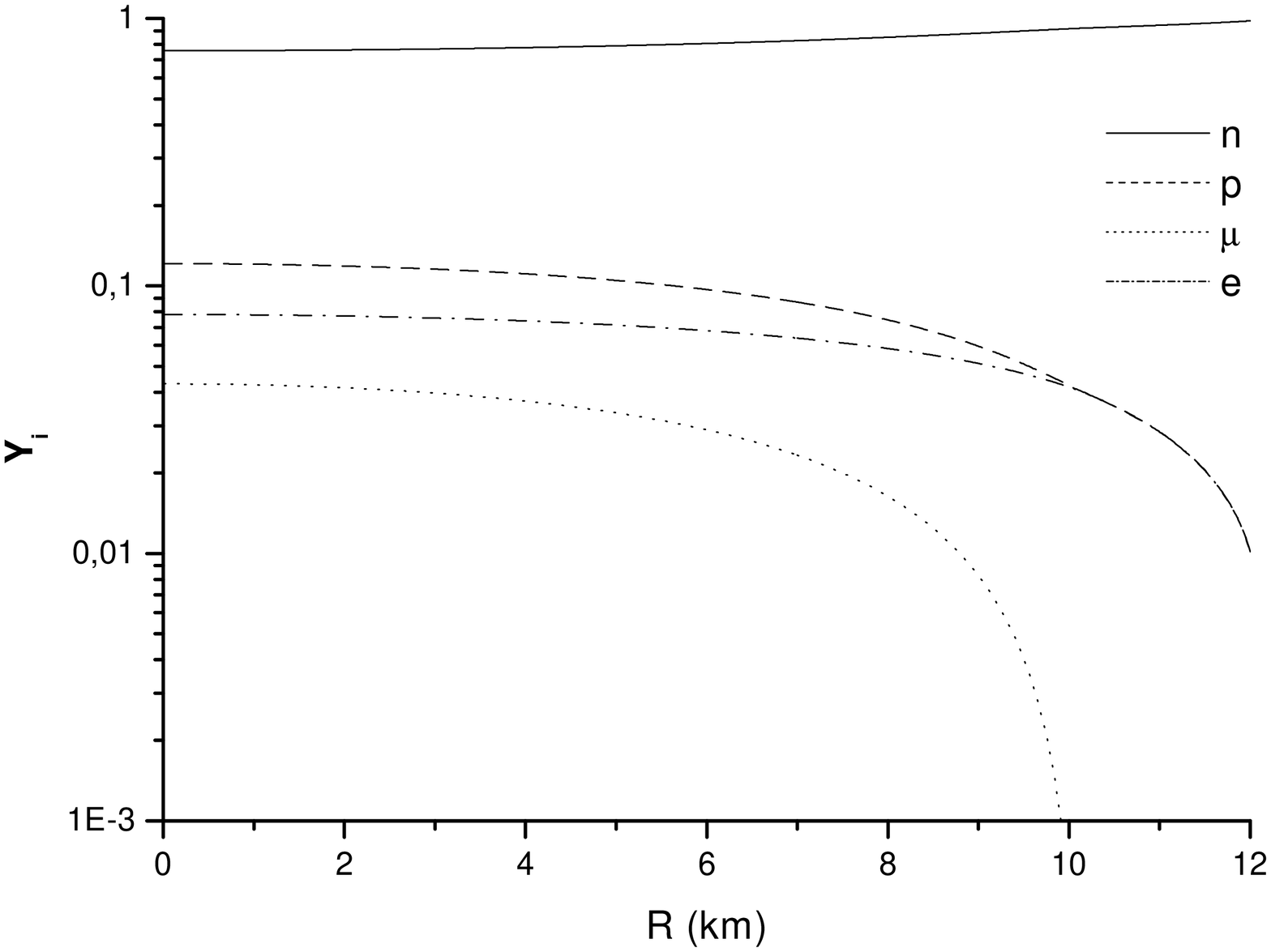}}}
\fbox{\subfigure[]
{\includegraphics[width=7.5cm]{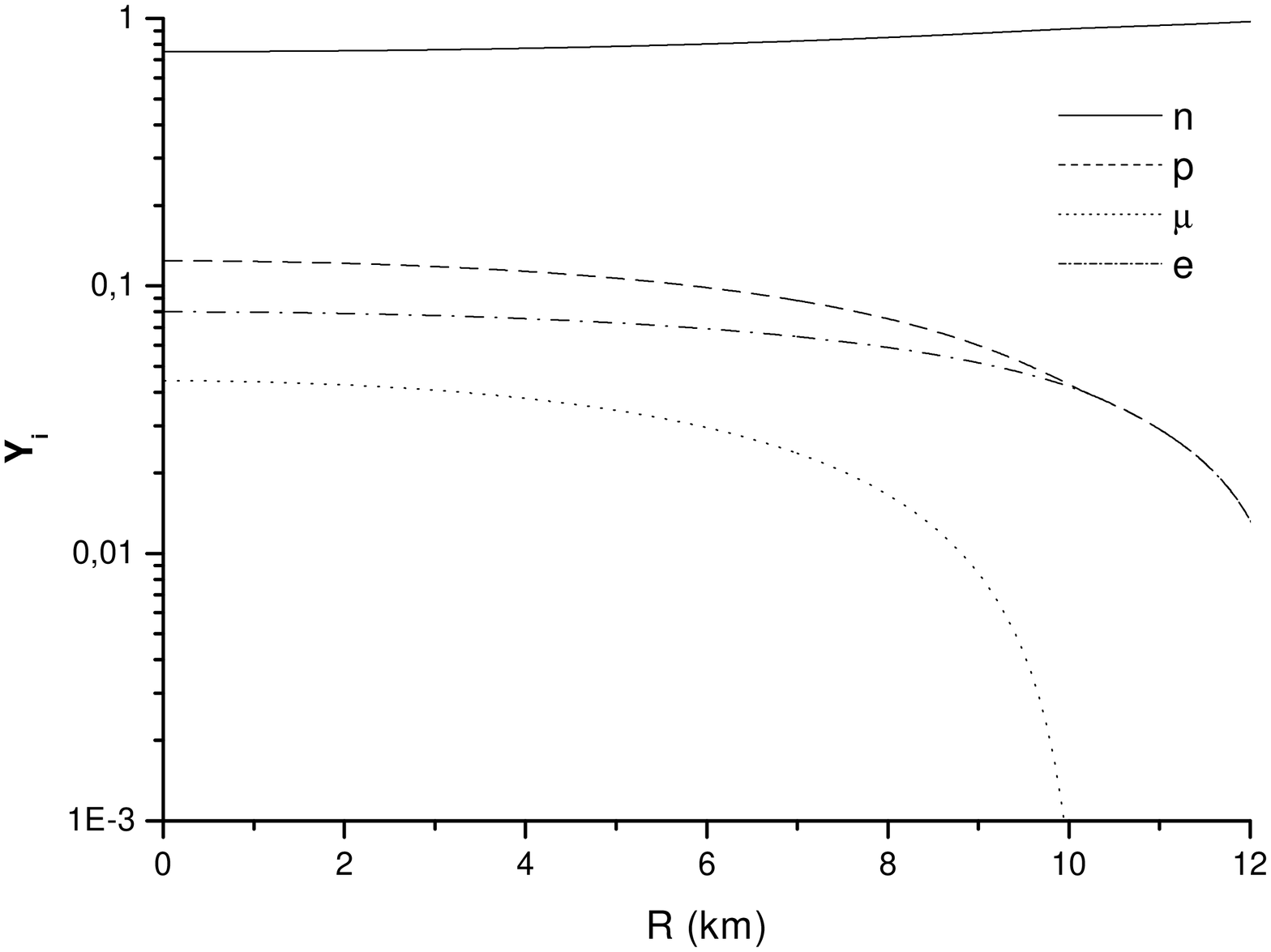}}}
\fbox{\subfigure[]
{\includegraphics[width=7.5cm]{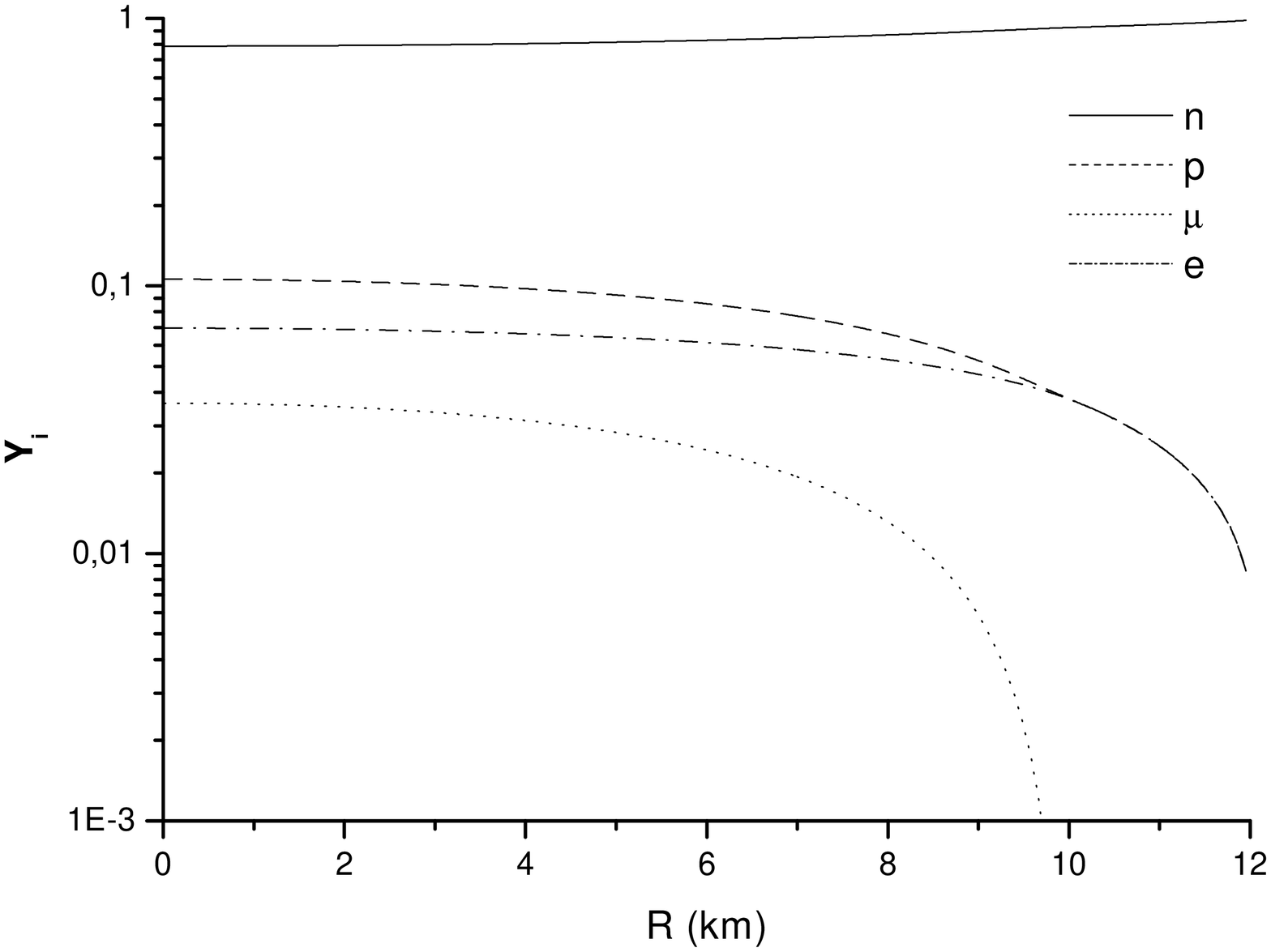}}}
\caption{The equilibrium compositions for the maximum radius
configuration as a function of the star radius. Panel (a) is for
the parameter set I whereas  (b) and (c) for parameter sets II and
 III.} \label{fig:partr}
\end{figure}
%%%%%%%%%%%%%%%%%%%%%%%%%%%%%%%%%%%%%%%%%%%%%%%%%%%%%%%%%%%%%%%
\begin{figure}
\fbox{\subfigure[]
{\includegraphics[width=7.5cm]{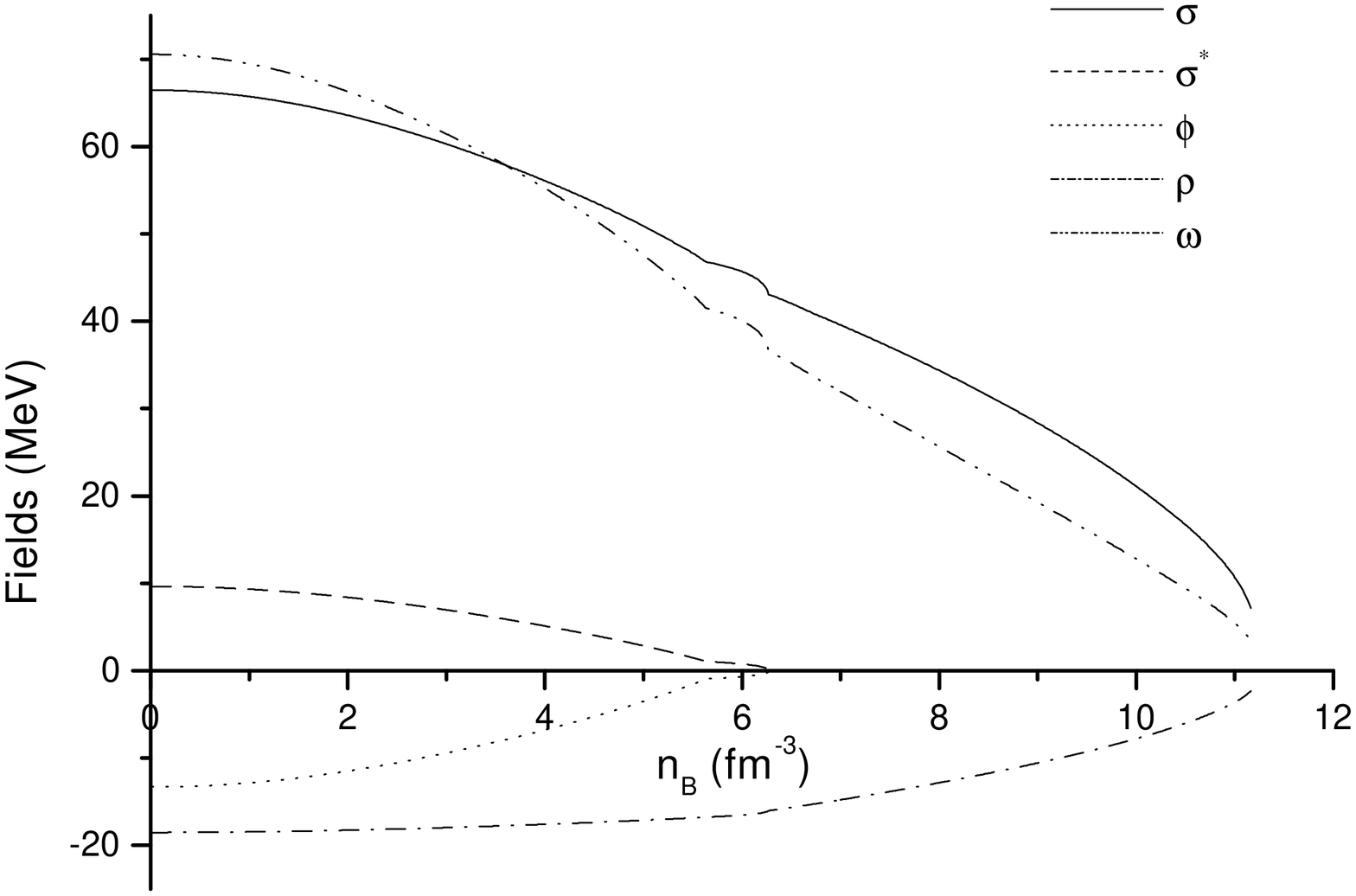}}}
\fbox{\subfigure[]
{\includegraphics[width=7.5cm]{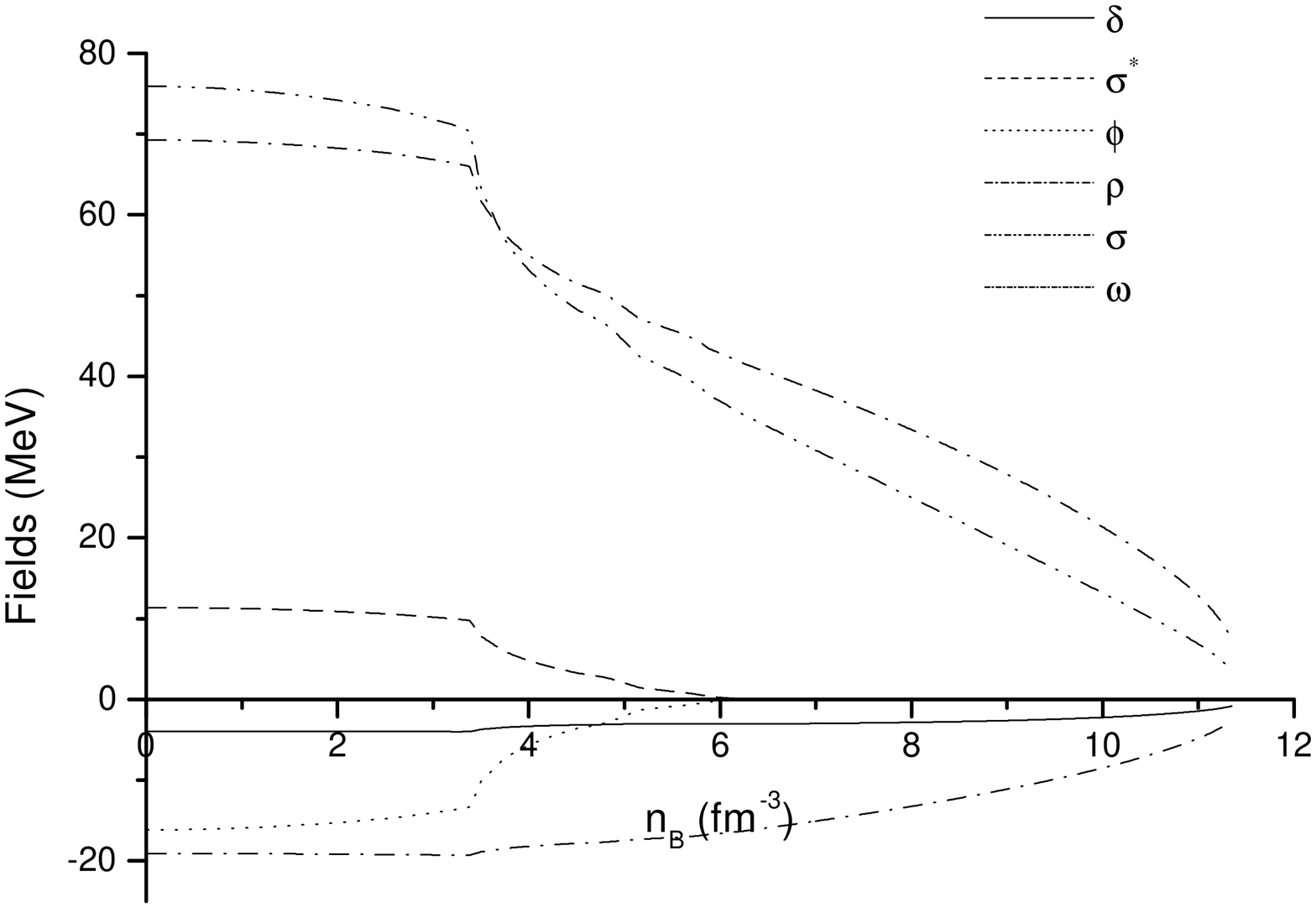}}}
\fbox{\subfigure[]
{\includegraphics[width=7.5cm]{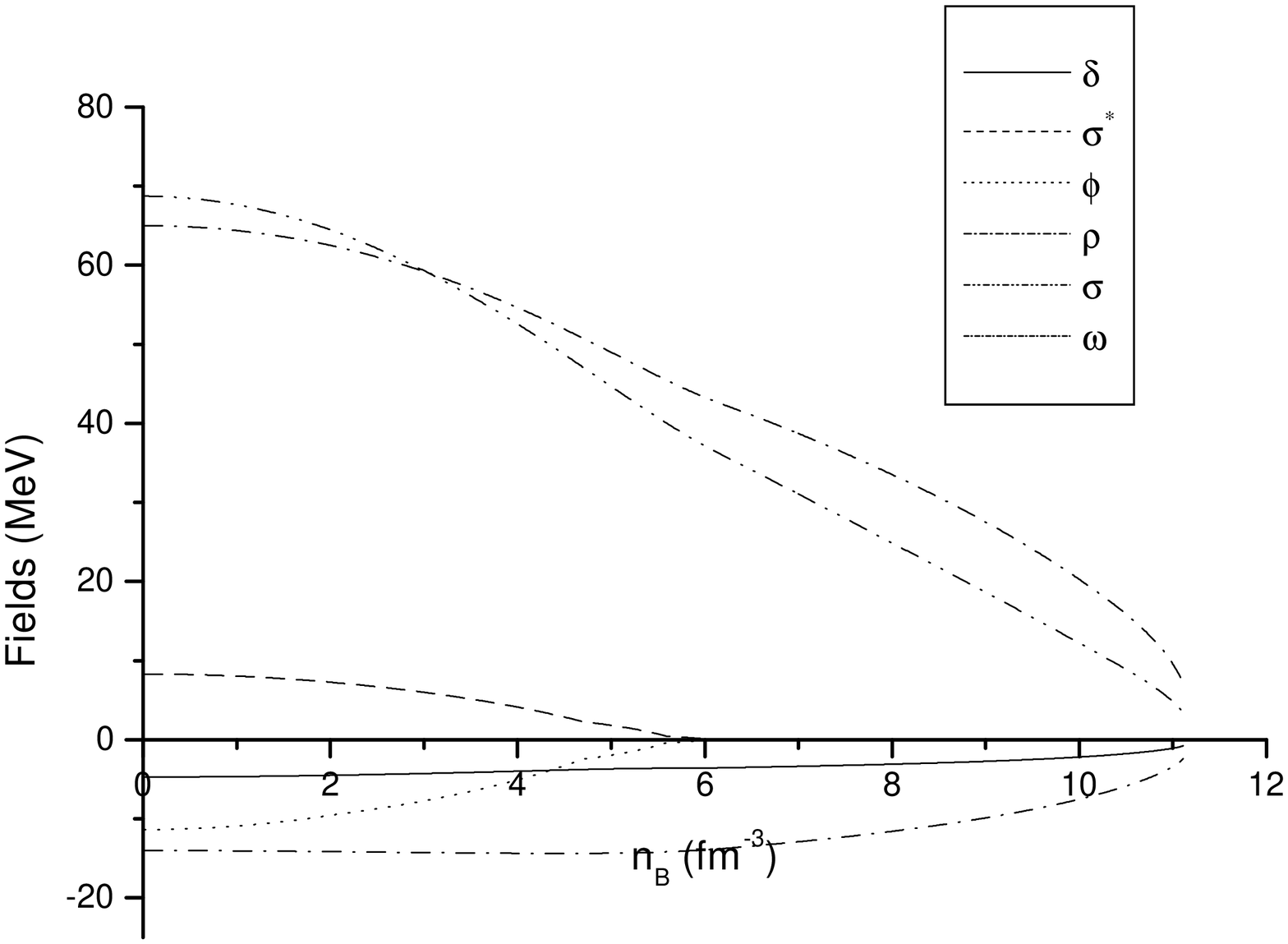}}}
\caption{The behavior of meson fields in the configuration with
maximum mass as functions of the star radius. Panels (a), (b) and
(c) are for parameter sets I, II and III respectively.}
\label{fig:fieldsm}
\end{figure}
%%%%%%%%%%%%%%%%%%%%%%%%%%%%%%%%%%%%%%%%%%%%%%%%%%%%%%%%%%%%%%%%
\clearpage
\begin{figure}
\fbox{\subfigure[]
{\includegraphics[width=7.5cm]{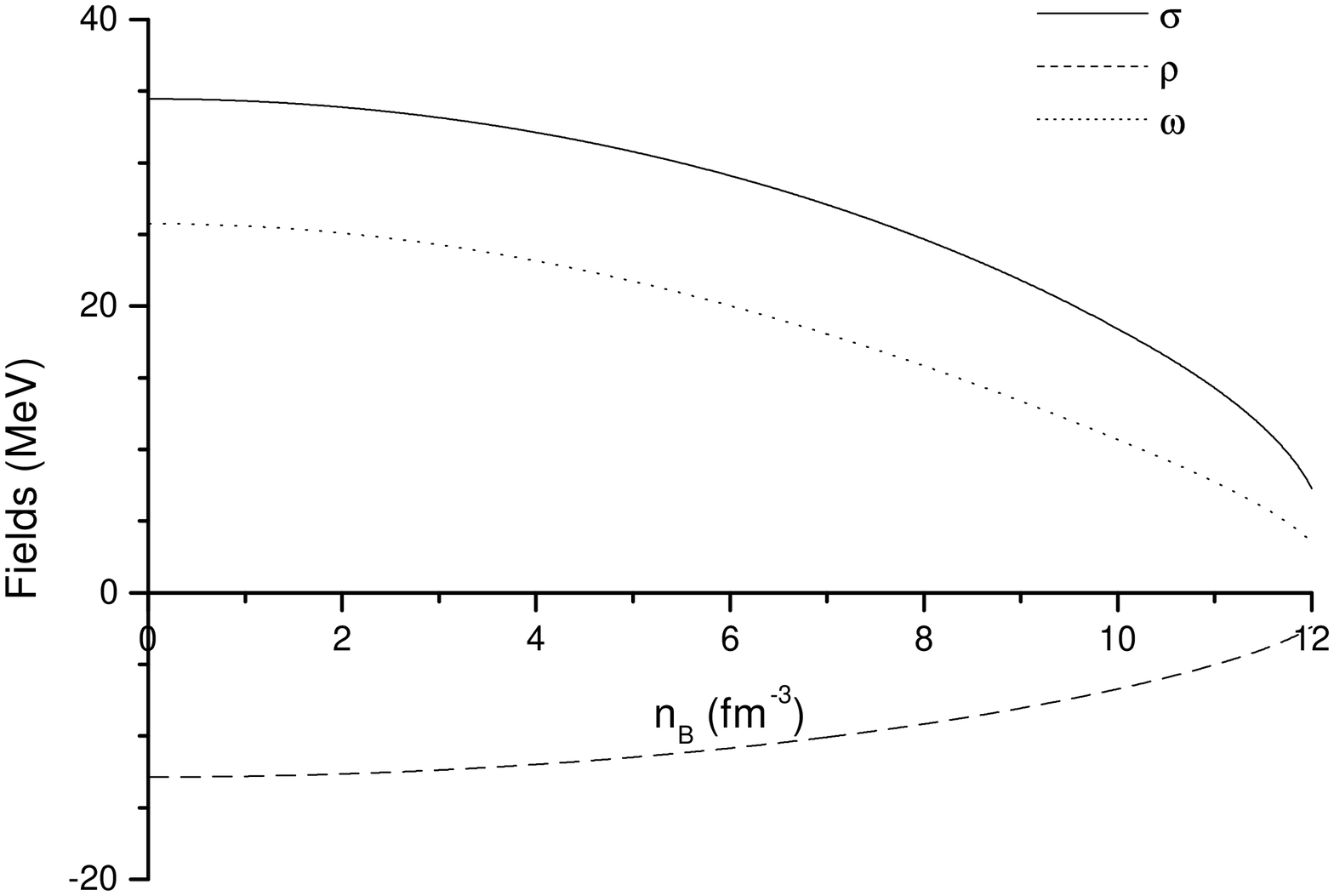}}}
\fbox{\subfigure[]
{\includegraphics[width=7.5cm]{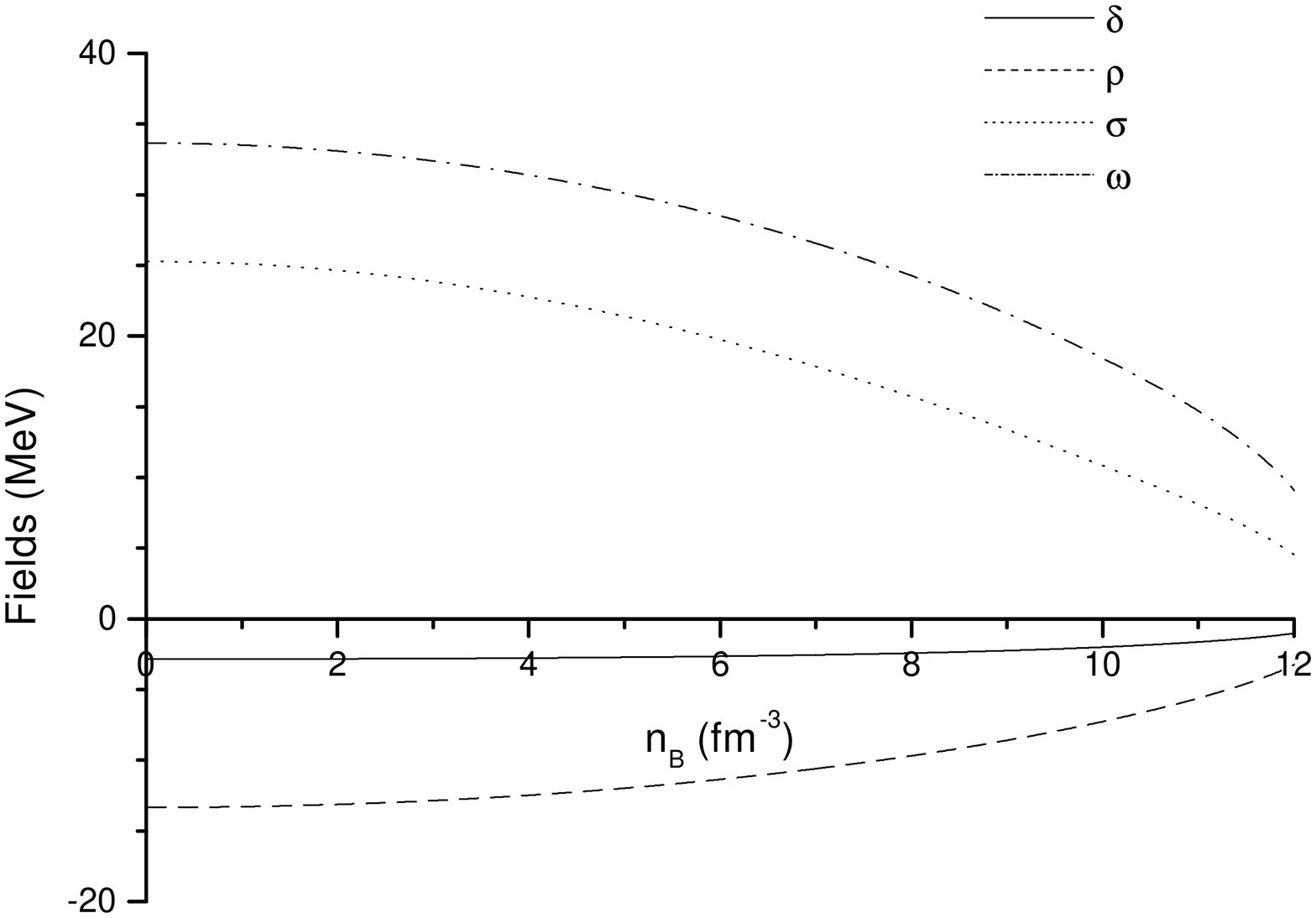}}}
\fbox{\subfigure[]
{\includegraphics[width=7.5cm]{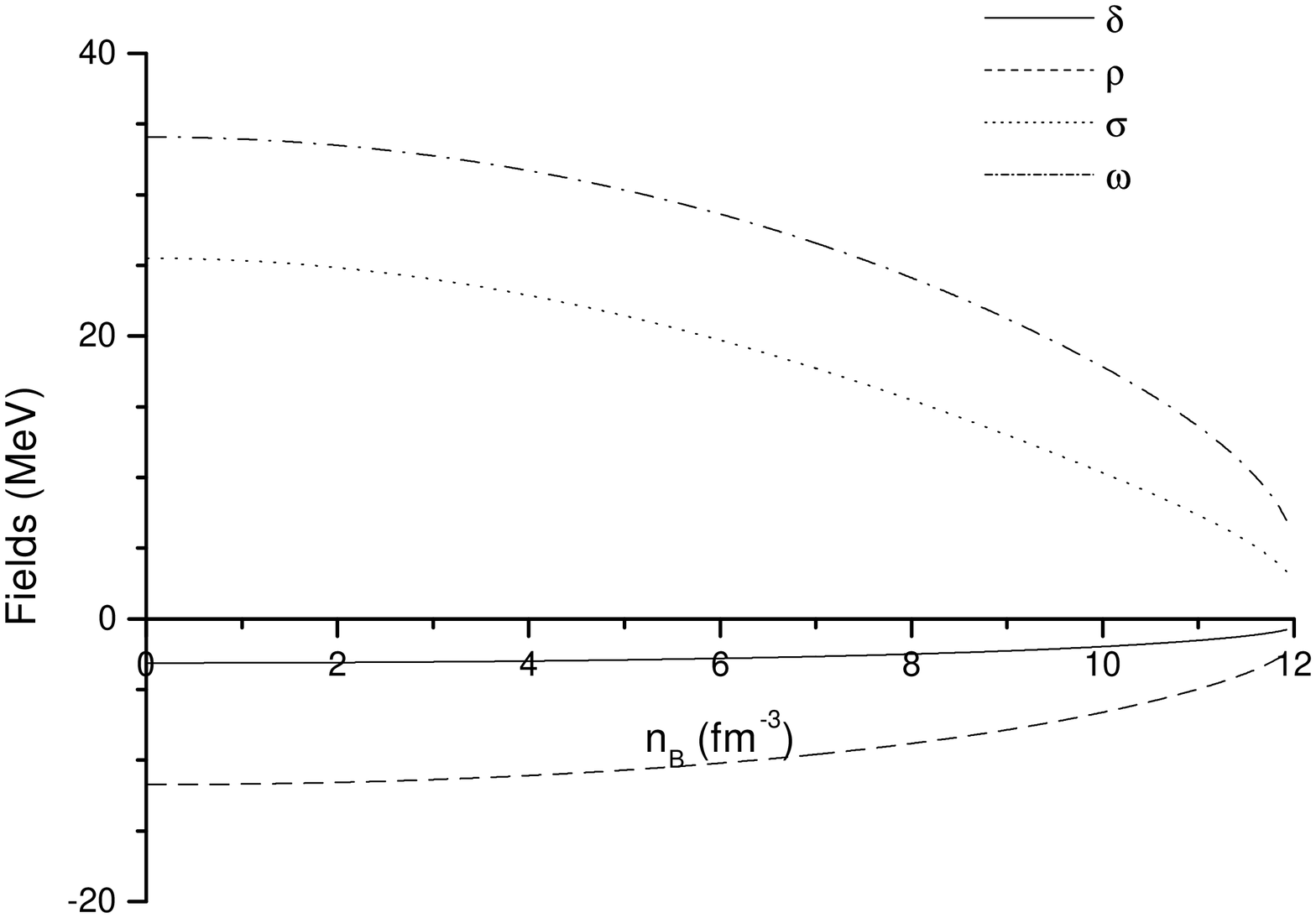}}}
\caption{The meson
fields in the neutron star with maximum radius as a function of
star radius.  The
panel (a) is for parameter set I, the panel (b) for parameter
set II the panel (c) for parameter set III.}
\label{fig:fieldsr}
\end{figure}
%%%%%%%%%%%%%%%%%%%%%%%%%%%%%%%%%%%%%%%%%%%%%%%%%%%%%%%%%%%%%%%%%
\section{Summary and conclusions}
In this paper the complete form of the equation of state of
hyperon matter has been obtained with the use of the derivative
coupling model in the framework of  an extended RMF theory which
besides hyperons and leptons includes  the extended meson sector
with additional  $\delta$ meson and hidden strange mesons
$\sigma^*$ and $\phi$. The model  considered is also supplemented
with nonlinear vector meson interactions. This enlargement alters
the symmetry properties of neutron star matter and through this
neutron stars parameters. The value of baryon effective masses
depend on the scalar meson condensates and at high densities when
hyperon species appear the possibility of negative nucleon masses
emerges. The derivative coupling model allows to avoid this
difficulty reproducing reasonable value of  baryon effective
masses for densities relevant for neutron stars. The inclusion of
 $\delta$ meson and nonlinear vector meson interactions
influences the chemical composition of a neutron star. This is
especially evident  comparing the effective baryon masses in the
density span $(3-4)\times \rho_0 $ and equilibrium compositions of
the star. For the third group of parameters the higher value of
asymmetry has been obtained. This changes the
properties of a neutron star diminishing the hyperon core extent.
The asymmetry of the system also influence the star radius, for
the third group of parameters the one can obtain the lower value
of the star radius. There is also possible a configuration
representing a star with lower densities (a maximum radius
configuration) which excluded the existence of a hyperon core. The
model considered excludes a large hyperon fraction which can be
connected with thermal properties of a hot neutron star. As the
most populated strange baryon is the $\Lambda$ hyperon it does not
reduce significantly number of leptons in the star interior and
thus the models calculated with the use of the parameter set I and
II  do not exclude rapid cooling rate of the star. This is even
more evident analyzing the proton fractions obtained for all
parameter groups. The equilibrium proton fraction is also
determined by the nuclear symmetry energy. The parameter sets I
and II permit higher values of proton fraction which is
indispensable for URCA processes to proceed. However, the
equilibrium proton fraction $Y_p$ is significantly reduced for the
third group of parameter.\\  Analyzing the gravitational binding
energy one can come to the conclusion that  configurations with
hyperons are energetically favorable than the one obtained with the
use of TM1 parameter set for higher densities.\\  All assumption which have been made
namely: the derivative coupling model  being connected  with the
higher effective baryon masses, the inclusion of $\delta$ meson
and nonlinear vector meson interactions, and the repulsive
nucleon-hyperon $\Sigma$ interaction lead to the neutron star
model with the value of maximum mass close to  $1.5 \ M_{\odot}$
with the reduced value of proton fraction and very compact hyperon
core. The calculation of the quark matter equation of state.
allows to construct the mass-radius relation for the quark star.
Comparing gravitational binding energies one can come to the
conclusion that the addition of hyperons to the model shifts the
stable hyperon matter configuration towards higher densities even
to the density range which is relevant for a quark star and at the
same time makes the existence of a pure quark star more
problematic.
%###################################################################

%#####################################################
%#####################################################
\end{document}